\documentclass[aps,notitlepage,nofootinbib,superscriptaddress]{revtex4-1}
\usepackage{amsmath,amsfonts,amssymb}
\usepackage{dsfont}
\usepackage{graphicx}
\usepackage{fancybox,bm}
\usepackage{empheq,bm}
\usepackage{color}

\newcommand{\Rb}{\mathbf{R}}
\newcommand{\Tb}{\mathbf{T}}
\newcommand{\rb}{\mathbf{r}}
\newcommand{\tb}{\mathbf{t}}
\newcommand{\kb}{\bm{k}}

\newcommand{\sigmab}{\bm{\sigma}}
\DeclareMathOperator{\Li}{Li}
\DeclareMathOperator{\tr}{tr}

\def\E{{\bm E}}
\def\D{{\bm D}}
\def\B{{\bm B}}
\def\S{\bm{ \mathcal S}}

\newcommand{\Ref}[1]{(\ref{#1})}
\newcommand{\be}{\begin{equation}}\newcommand{\ee}{\end{equation}}
\newcommand{\II}{I\!I}
\newcommand{\XY}{\mathrm{XY}}
\newcommand{\EM}{\mathrm{EM}}
\newcommand{\si}{\sigma}

\newcommand{\tk}{\tilde{k}}
\newcommand{\al}{\alpha}
\newcommand{\ep}{\epsilon}

\def\({\left(}
\def\){\right)}
\def\[{\left[}
\def\]{\right]}

\newcommand{\ii}{\mathrm{i}}
\newcommand{\wb}{\mathbf{w}}
\newcommand{\nxi}{\zeta}

\usepackage{color} \definecolor{darkgreen}{rgb}{0,.5,0}
\usepackage[colorlinks,filecolor=blue,citecolor=darkgreen,unicode]{hyperref}

\begin{document}
\title{The quest for Casimir repulsion between Chern-Simons surfaces}
\author{Ignat Fialkovsky}
\email{ifialk@gmail.com}
\affiliation{CMCC, Universidade Federal do ABC, Avenida dos Estados 5001, CEP 09210-580, SP, Brazil}
\author{Nail Khusnutdinov}
\email{nail.khusnutdinov@gmail.com}
\affiliation{CMCC, Universidade Federal do ABC, Avenida dos Estados 5001, CEP 09210-580, SP, Brazil}
\affiliation{Institute of Physics, Kazan Federal University, Kremlevskaya 18, Kazan, 420008, Russia}
\author{Dmitri Vassilevich}
\email{dvassil@gmail.com}
\affiliation{CMCC, Universidade Federal do ABC, Avenida dos Estados 5001, CEP 09210-580, SP, Brazil}
\affiliation{Department of Physics, Tomsk State University, 634050 Tomsk, Russia}

\begin{abstract}
In this paper we critically reconsider the Casimir repulsion between surfaces that carry the Chern-Simons interaction (corresponding to the Hall type conductivity). We present a derivation of the Lifshitz formula valid for arbitrary planar geometries and discuss its properties. This analysis allows us to resolve some contradictions in the previous literature. We compute the Casimir energy for two surfaces that have constant longitudinal and Hall conductivities. The repulsion is possible only if both surfaces have Hall conductivities of the same sign. However, there is a critical value of the longitudinal conductivity above which the repulsion disappears. We also consider a model where both parity odd and parity even terms in the conductivity are produced by the polarization tensor of surface modes. In contrast to the previous publications \cite{Chen:2011,Chen:2012ds}, we include the parity anomaly term. This term ensures that the conductivities vanish for infinitely massive surface modes. We find that at least for a single mode regardless of the sign and value of its mass, there is no Casimir repulsion.
\end{abstract}

\maketitle
\section{Introduction}
The Casimir effect is one of the best studied manifestations of the non-trivial quantum vacuum structure \cite{Bordag:2001qi,Milton:2004ya,Bordag:2009zzd}. It became a valuable tool for studying new materials \cite{Woods:2015pla}. One of the most exciting recent discoveries is the possibility of a Casimir repulsion between the surfaces that carry Chern-Simons actions.

As far as we know, for the first time boundary surfaces supplied with a Chern-Simons term were considered in the context of Quantum Field Theory in the papers \cite{Elizalde:1998ha,Bordag:1999ux} where the heat kernel coefficients and the Casimir energy were computed. In the paper \cite{Bordag:1999ux} a Casimir repulsion was observed. Those papers, however, considered rigid non-penetrable boundary conditions modified by the Chern-Simons term rather than $P$-odd (Hall) conductivities on the interface surfaces. The latter is the situation to be addressed in the present work. The Casimir interaction of interfaces with a constant Hall conductivity was computed in \cite{Markov:2006js} by evaluating the corresponding vacuum energy densities with Quantum Field Theory methods. Recently, this result was reconfirmed with the use of Lifshitz formula \cite{Marachevsky:2017tdo}. In the paper \cite{Fosco:2016rwf} also some momentum-dependent form-factors in both Hall and longitudinal conductivities were added. Some general aspects of gauge invariance of the Casimir energy in the presence of a parity breaking $\theta$-term were considered in \cite{Canfora:2011fd}. The Hall conductivity appears if a two-dimensional electron system is placed in a strong perpendicular magnetic field. The Casimir force between two such systems was calculated in \cite{Tse:2012}. 

Quite naturally, there has been much activity in this direction in the context of new materials. The Casimir force in staggered $2D$ graphene family materials (silecene, germanene and stanene) was investigated in Ref. \cite{Rodriguez-Lopez:2017:cfptitgf}. Due to the topological properties of these materials, repulsive and quantized Casimir interactions become possible. The papers \cite{Grushin:2011b,Grushin:2011,RodriguezLopez:2011dq,Nie:2013} computed the Casimir interaction between Topological Insulators (TIs) that were modeled by a dielectric bulk carrying a constant Hall conductivity on the surface. The Casimir stress on a TI between two metallic plates was investigated in \cite{Martin-Ruiz:2016qms}. In the works \cite{Chen:2011,Chen:2012ds} the surface conductivity of TIs was produced by the one-loop effects of finite gap surface fermions. The TIs discussed above are also called axionic TIs. The Casimir force between photonic TIs was computed in \cite{Fuchs:2017uyg}. These latter materials do not show a Chern-Simons term on the surface and thus do not fit into the subject of present work. The Casimir force between Chern insulators was computed in \cite{Rodriguez-Lopez:2013pza}. Finally, we like to mention the work \cite{Wilson:2015} dedicated to the Casimir interaction of Weyl semimetals even though the system considered there is quite different -- the Chern-Simons interaction is present in the bulk of Weyl semimetals. 

Probably the most interesting result of the papers that we have mentioned above is the possibility of Casimir repulsion\footnote{The literature on Casimir repulsion as such is very large. Some overview can be found in \cite{Milton:2012ti}.}  in some ranges of the parameters used. However, looking more attentively at the liteature, one observes that this repulsion was observed in \cite{Markov:2006js,Marachevsky:2017tdo,Tse:2012,Grushin:2011b,Grushin:2011,RodriguezLopez:2011dq} for the same sign of the Hall conductivity on interacting surfaces, while in \cite{Chen:2011,Chen:2012ds,Rodriguez-Lopez:2013pza} -- for opposite signs of the conductivity. Postponing a detailed analysis of \cite{Chen:2011,Chen:2012ds,Rodriguez-Lopez:2013pza} to a due time (see Secs. \ref{sec:part} and \ref{sub:ano} below), we note that all these papers used the Lifshitz formula in TE-TM basis to evaluate the Casimir energy.

The Lifshitz formula (see, e.g., \cite{Bordag:2009zzd}) relates the Casimir energy to an integral of the scattering data. Undoubtedly, this formula is the main tool for Casimir calculations with plane boundaries.  All information on two interacting bodies, $I$ and $\II$, is encoded in the product of two reflection matrices $\rb'_I\rb^{\phantom{\prime}}_{\II}$, where the prime means, roughly speaking, the inversion of scattering direction. This prime is omitted in some papers without warning if its presence or absence does not alter the result. The authors of \cite{Chen:2011,Chen:2012ds,Rodriguez-Lopez:2013pza} took their Lifshitz formulas from one of such papers, though the prime does influence the result for Casimir energy in the case of Topological Insulators. Even though this seems to be a plausible explanation for the discrepancies, one cannot guarantee that this is the end of the story and is urged to rederive the Lifshitz formula. The history of \emph{derivations} of Lifshitz formula is long and dramatic, see \cite{Nesterenko:2011ab} for an extensive list of references. All derivations were performed under some explicit or implicit restrictions on the properties of interacting systems. Strictly speaking, non is fully suitable for a description of the Casimir interaction between the surfaces with Hall type conductivities.

In our derivation of the Lifshitz formula, we start with an expression for the Casimir energy as a sum over the photon normal frequencies expressed through the scattering matrix, transform the sum to a contour integral and rotate the contour. In its spirit, our approach reminds that of \cite{Bordag1995,Genet:2003zz,Lambrecht:2006} though we do not impose any restrictions on the scattering amplitudes except for the symmetry with respect to translations along some plane. We are however somewhat sloppy with the analytic properties of integrals since this aspect has been sorted out completely in \cite{Nesterenko:2011ab}. Our rather general consideration confirms the Lifshitz formula (with a prime in $\rb'_{I}$, of course). We also derive general and quite useful formulas for the transition between the TE-TM basis and the basis XY of linear polarizations. All this is done in Sec.~\ref{sec:Cas}.

In Sec. \ref{sec:refl} we proceed with a bit more specific case of a surface with arbitrary momentum-dependent conductivity and a isotropic substrate characterized by a permittivity $\epsilon$ and a permeability $\mu$. For this case we derive the reflection matrix and study the question: What is the price for deleting the prime in $\rb'$? It appears, that in the XY basis $\rb$ and $\rb'$ coincide, which is not true in the TE-TM basis. However, if the conductivity tensor satisfies some additional condition, passing from $\rb'$ to $\rb$ in TE-TM basis is essentially equivalent to the inversion of sign in front of the Hall conductivity. This condition is satisfied in the cases considered in \cite{Chen:2011,Chen:2012ds,Rodriguez-Lopez:2013pza}. 

In Sec. \ref{sec:part} we consider the interaction of two planes with constant conductivities in the vacuum. All formulas greatly simplify in this case. We analyze the interaction (i) of two identical surfaces with constant conductivities, (ii) of two surfaces with identical longitudinal conductivities and Hall conductivities of opposite sign, (iii) interaction of a surface with both Hall and longitudinal conductivities with an ideal conductor. Only in the case (i) there can be an interval of values of the Hall conductivity corresponding  to the Casimir repulsion. This interval disappears if the longitudinal conductivity is larger than some critical value.

In Sec.\ \ref{sub:ano} we study a more sophisticated model where the surface conductivity is expressed through the one-loop polarization tensor of massive boundary modes. This is essentially the model considered in \cite{Chen:2011,Chen:2012ds}, but with two modifications. One refers to the issue of prime in $\rb'$, which is very easy to correct. The other is more important. The expression for polarization tensor used in \cite{Chen:2011,Chen:2012ds} does not include the parity anomaly contribution \cite{Niemi:1983rq,Redlich:1983dv}. Among other properties, the parity anomaly term ensures that infinitely massive particles give no contribution to the conductivity. We find this property very natural from the physical point of view. We compute the Casimir energy with the parity anomaly term included and find that the Casimir repulsion is not possible in this case for a single boundary mode. Moreover, the Hall conductivity gives just a tiny contribution to the Casimir interaction.

\section{Casimir energy in planar geometries}\label{sec:Cas}
In this section we derive general formulas for the Casimir energy for planar geometries that are suitable for computations if the matching conditions on some of the plains contain a $P$-odd (Hall) part. The main peculiarity of this case as compared to a more common $P$-even scattering is the absence of some symmetry properties of scattering matrix. This will be the main concern here. We shall pay less attention to some other aspects, as regularization or absorption, for example, as these aspects have been thoroughly studied in the literature. 

\subsection{Scattering matrix and the spectrum}\label{sec:scat}
\begin{figure}
	\includegraphics[width=4.8cm]{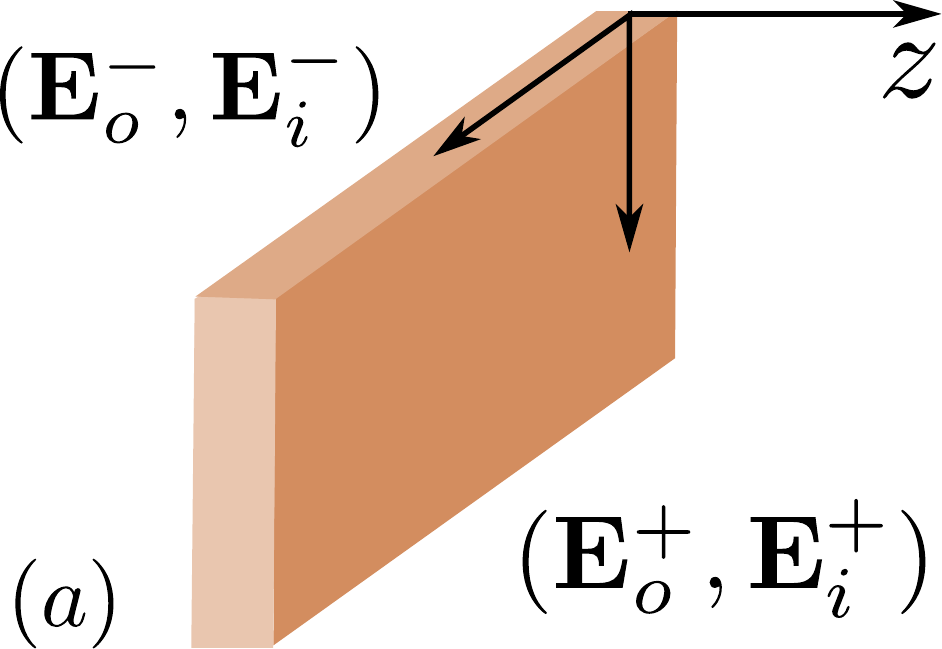}\hspace{6em}\includegraphics[width=7.5cm]{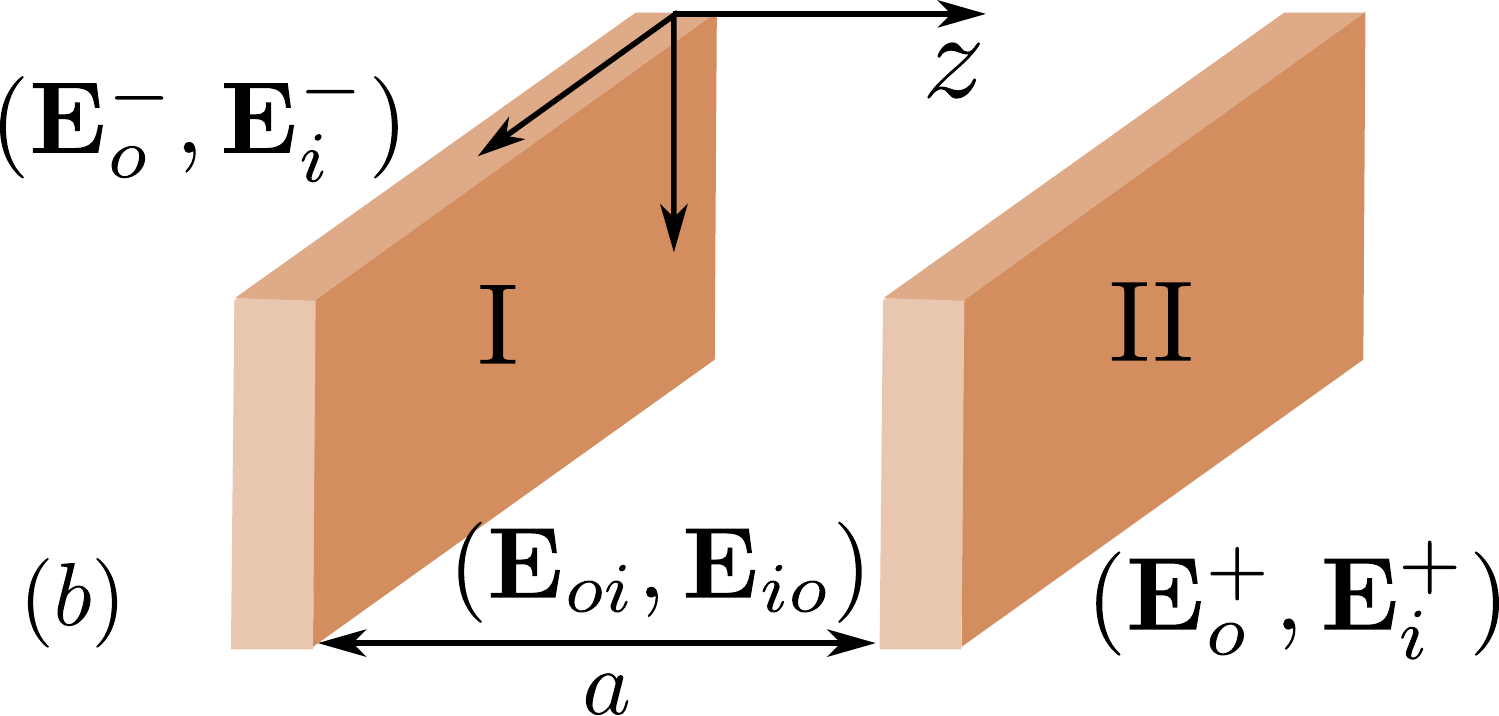}
	\caption{The geometries discussed in Sections \ref{sec:scat} and \ref{sec:two}: a planar system between two asymptotic regions at $z\to\pm \infty$ (panel a) and two planar systems separated by a distance $a$ (panel $b$). }\label{fig:1n}
\end{figure}

Here we consider the scattering in plain geometries that have two well-defined asymptotic regions at $x^3=z\to \pm\infty$ and possess translational symmetries in the orthogonal directions, see Fig.\ \ref{fig:1n}(a). Through the Maxwell equations, all components of electromagnetic field in both asymptotic regions may be expressed through components of the electric field $(E_1,E_2)^T=(E_x,E_y)^T$ parallel to the plain. (Here the superscript $T$ means the matrix transposition.) Thus, it is enough to consider the scattering of these components only. It is convenient to combine the fields $\E^-$ and $\E^+$ corresponding to large negative and large positive values of $z$, respectively, into a single four-component object
\begin{equation}
\E=\(\E^-,\E^+\)^T
  =\(\E^-_i e^{\ii k_3z}+\E^-_o e^{-\ii k_3z},\E^+_i e^{-\ii k_3z}+\E^+_o e^{\ii k_3z}\)^T,
\label{Eio}
\end{equation}
where the subscripts $i$ and $o$ are used to denote incoming and outgoing waves. Here and in what follows we do not write explicitly the common oscillating factor $e^{ik_0x^0+ik_1x^1+ik_2x^2}$.  Then the scattering matrix $\S$ is defined as
\begin{equation}
\(\E^{-}_o,\E^{+}_o\)^T = 
		\S \(\E^{-}_i,\E^{+}_i\)^T\,.\label{Smat}
\end{equation}
More explicitly,
\begin{equation}
\( \begin{array}{c} \E^-_o \\ \E^+_o  \end{array} \) =
\( \begin{array}{cc} \Rb & \Tb' \\ \Tb & \Rb'  \end{array}\)
\( \begin{array}{c}  \E^-_i \\ \E^+_i  \end{array}\) \,,
\label{Smat2}
\end{equation}
where $\Rb$, $\Rb'$, $\Tb$ and $\Tb'$ are $2\times 2$ reflection and transmission matrices.

Further on, the whole field can be written as a function of the incoming wave
\begin{equation}
\E= (\mathcal{K}+\mathcal{K}^* \S ) \E_i
\end{equation}
with $\E_i\equiv (\E_i^-,\E_i^+)^T$ and 
\begin{equation}
\mathcal{K} = \( \begin{array}{cc} e^{\ii k_3z} {\bf 1}_2 & 0 \\ 0 & e^{-\ii k_3z} {\bf 1}_2 \end{array}\)\,.
\end{equation}
${\bf 1}_2$ is a unit $2\times 2$ matrix.

To relate the Casimir energy to the scattering data we follow the method of Ref.\ \cite{Bordag1995}.
We introduce auxiliary boundaries located at $z=\pm L$. Later, we shall take the limit $L\to\infty$. We impose the perfect conductor boundary conditions at these boundaries,
\begin{equation}
\E^-(z=-L) =0,\qquad 
  \E^+(z=+L) =0 \,.
\end{equation}
After some algebra these conditions may be combined into a single condition
\begin{equation}
\(
  \begin{array}{cc}
   e^{-\ii k_3 L} {\bf 1}_2+ e^{\ii k_3 L} \Rb  &  e^{\ii k_3 L} \Tb'\\
   e^{\ii k_3 L} \Tb &  e^{-\ii k_3 L}{\bf 1}_2+ e^{\ii k_3 L} \Rb'\\
  \end{array}
  \)
  \E_i
=\(e^{-\ii k_3 L} {\bf 1}_4 +  e^{\ii k_3 L}  \S  \)  \E_i
  =0\,,\label{eq:nMod}
\end{equation}
which has to be considered as an equation to define the incident waves. The equation has a solition if and and only if
\begin{equation}
f(k)\equiv \det\(e^{-\ii k_3 L} {\bf 1}_4 +  e^{\ii k_3 L}  \S  \)  =0 \,.\label{eq:spec}
\end{equation}
This condition defines the spectrum of allowed values of the transverse momentum $k_3$.

The vacuum energy $\mathcal{E}$ is formally defined as one half the sum over the eigenfrequencies of the normal modes of electromagnetic field. In our case, it also includes the integration over momenta $k_1$ and $k_2$, so that
\begin{equation}
\mathcal{E}=\frac12\sum_J \int \frac{d^2 k}{(2\pi)^2} \sqrt{k_1^2+k_2^2+k_{3,J}^2(k_1,k_2)}\,\,,
\label{eq:mE}
\end{equation}
where $k_{3,J}(k_1,k_2)$ denotes distinct solutions of (\ref{eq:spec}) for given $k_1$, $k_2$, numbered by $J$.  The expression (\ref{eq:mE}) is divergent and, in principle, has to be regularized. Some regularization methods for the vacuum energy are described in \cite{Bordag:2001qi}. Since the Casimir force between distinct bodies is finite, we skip this step and do not mention any explicit regularization.

\begin{figure}
	\includegraphics[width=6cm]{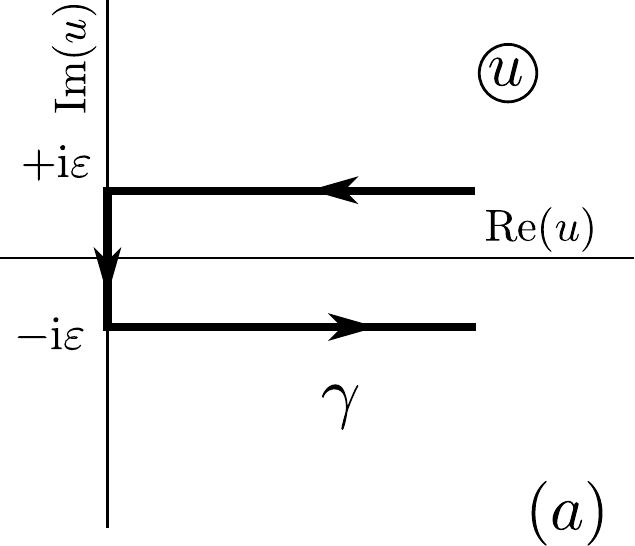}\hspace{6ex}\includegraphics[width=6cm]{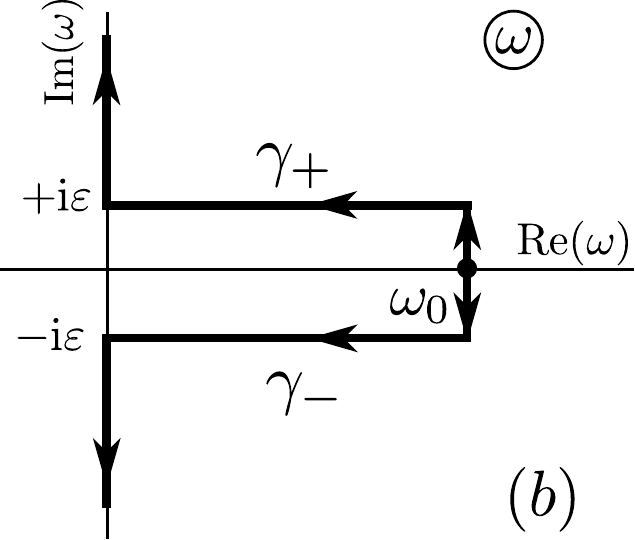}
	\caption{The contours in complex plain used for integration  (a) in Eq.\ (\ref{Eint2}) and (b) in Eq.\ (\ref{Lif3}).  }\label{fig:2n}
\end{figure}

As another simplification we neglect the effects of absorption and possible plasmonic modes. The necessary modifications in the derivation of the Lifshitz formula for Casimir energy have been studied in all detail in Ref.\  \cite{Nesterenko:2011ab}. We have nothing to add to this paper. Thus, under this assumption, the spectrum of $k_3$ is real and positive. We may represent (\ref{eq:mE}) as a contour integral
\begin{equation}
\mathcal{E}=\frac1{4 \pi \ii}  \int \frac{d^2 k}{(2\pi)^2}\oint_\gamma du\, \sqrt{u^2 +k_1^2+k_2^2} \frac{d}{du} \ln f(k_1,k_2, u)
\end{equation}
with $\gamma = (+\infty +\ii\varepsilon, 0+\ii\varepsilon)\cup (0+\ii\varepsilon,0-\ii\varepsilon)\cup (0-\ii\varepsilon,\infty -\ii\varepsilon)$ running in the anti-clockwise direction around the positive real axis, see Fig.\ \ref{fig:2n}(a). After integration by parts the integral becomes
\begin{equation}
\mathcal{E}=- \frac1{4 \pi \ii}  \int \frac{d^2 k}{(2\pi)^2}\oint_\gamma \frac{u du}{\sqrt{u^2 +k_1^2+k_2^2}} \ln f(k_1,k_2, u)\,.\label{Eint2}
\end{equation}

Let us consider the upper part of the contour $\gamma$, $u=k_3+i\varepsilon$. For large values of $L$
\begin{equation}
\ln f(k_1,k_2,u) \simeq -4i k_3 L+4 \varepsilon L + e^{2 i k_3 L-2 \varepsilon L}   \mathrm{tr}\, \S \,.
\end{equation}
The first and second terms on the right hand side do not depend on the properties of the system and can thus be neglected. The last term is exponentially small at $L\to \infty$. At the lower part of the contour, taking $u=k_3-\ii\varepsilon$, we have
\begin{equation}
\ln f(k_1,k_2,u) = -4\ii k_3 L+4 \varepsilon L +\ln \det \bigl(e^{-\ii k_3L -\varepsilon L}{\bf 1}_4 +\S \bigr) \,.
\end{equation}
The first two terms can be again neglected. In the last term, the corrections to $\S$ are exponentially small. Thus, $\ln\det (\S)$ is the only term which survives in the $L\to \infty $ limit. Next, we take the limit $\varepsilon\to 0$ to obtain
\begin{equation}
\mathcal{E}= - \frac1{4 \pi \ii}  \int \frac{d^2 k}{(2\pi)^2} 
	\int_0^\infty \ln \det \S\(k_1, k_2, k_3\) \frac{k_3 dk_3}{\sqrt{k_1^2+k_2^2+k_3^2}} \,.
	\label{ii_Lif1}
\end{equation}
This is a known formula. With some variations one can find it, for instance, in \cite{Bordag1995,Reynaud:2010vr,Ingold:2014uoa}.

The scattering matrix (\ref{Smat}) is defined for positive values of $k_3$ only. Inversion of the sign in front of $k_3$ just means the interchange of incoming and outgoing modes. This allows us to define $\S$ for negative $k_3$ by the relation
\begin{equation}
\S(-k_3)\equiv \S^{-1}(k_3)\,.\label{Smin}
\end{equation}
The identity $\S(-k_3)\S(k_3)={\bf 1}_4$ yields for reflection and transmission matrices
\begin{eqnarray}
\Tb^{-1}(-k_3) &=& \Tb' - \Rb \Tb^{-1} \Rb',\nonumber\\
\Rb'^{-1}(-k_3) &=& \Rb' - \Tb \Rb^{-1} \Tb',\nonumber\\
\Tb'^{-1}(-k_3) &=& \Tb - \Rb' \Tb'^{-1} \Rb,\nonumber\\
\Rb^{-1}(-k_3) &=& \Rb - \Tb' \Rb'^{-1} \Tb.\label{TRmTR}
\end{eqnarray} 
On the right hand sides of these relations all quantites are taken at $+k_3$. Thus
\begin{equation}
\det \S = \det
\begin{pmatrix}
\Rb & \Tb'\\
\Tb & \Rb'
\end{pmatrix}
= \det \Rb \det \left(\Rb' - \Tb \Rb^{-1} \Tb' \right) = \frac{\det \Rb(+k_3)}{\det \Rb'(-k_3)}.
\label{SRR}
\end{equation}

\subsection{Interaction of two planar systems}\label{sec:two}
Let us now consider two systems having a planar symmetry and separated by a vacuum gap between $z=0$ and $z=a$ as on Fig\ \ref{fig:1n}(b). The modes $\E_o^-$ and $\E_i^-$ for the left system, and $\E_o^+$ and $\E_i^+$ for the right system still play the same role of asymptotic modes for the whole system. The mode $\E_o^+$ (resp., $\E_i^+$) for the left system coincides with the mode $\E_i^-$ (resp., $\E_o^-$) for the right system. We denote these modes by $\E_{oi}$ and $\E_{io}$, respectively. Each of the systems is characterized by its own scattering matrix, so that
\begin{equation}
\begin{pmatrix} \E_{o}^- \\ \E_{oi} \end{pmatrix}=
\begin{pmatrix}
\rb_0 & \tb'_0 \\
\tb_0 & \rb'_0
\end{pmatrix}
\begin{pmatrix} \E_{i}^- \\ \E_{io} \end{pmatrix}\,,\qquad
\begin{pmatrix} \E_{io} \\ \E_0^+ \end{pmatrix} =
\begin{pmatrix}
\rb_a & \tb'_a \\
\tb_a & \rb'_a
\end{pmatrix}
\begin{pmatrix} \E_{oi} \\ \E_i^+ \end{pmatrix}\,.
\end{equation}
From these equations one can express $\E_{oi}$ and $\E_{oi}$ as
\begin{eqnarray}
&&\E_{io}=(1-\rb_a\rb_0')^{-1} \bigl( \rb_a\tb_0\E_i^- +\tb_a'\E_i^+\bigr),\nonumber\\
&&\E_{oi}=(1-\rb_0'\rb_a)^{-1} \bigl( \tb_0\E_i^- + \rb_0'\tb_a'\E_i^+ \bigr), \nonumber
\end{eqnarray}
and find the total scattering matrix
\begin{equation}
\S = \begin{pmatrix}\label{eq:S0a}
\rb_0 + \tb'_0 \left(1 - \rb_a \rb'_0\right)^{-1} \rb_a\tb_0 & \tb'_0 \left(1 - \rb_a \rb'_0\right)^{-1} \tb'_a \\
\tb_a \left(1 - \rb'_0 \rb_a\right)^{-1} \tb_0 & \rb'_a + \tb_a \left(1 - \rb'_0 \rb_a\right)^{-1} \rb'_0 \tb'_a
\end{pmatrix}.
\end{equation}
This implies that $\Rb= \rb_0 + \tb'_0 \left(1 - \rb_a \rb'_0\right)^{-1} \rb_a\tb_0$ and $\Rb' = \rb'_a + \tb_a \left(1 - \rb'_0 \rb_a\right)^{-1} \rb'_0 \tb'_a$. One can easily derive the identity
\begin{equation}
\rb{}^{-1}_a \left(1 - \rb{}_a \rb{}'_0\right) \tb{}'^{-1}_0 \Rb \rb{}^{-1}_0 \tb{}'_0 = -\left(\rb'_0 - \tb_0 \rb^{-1}_0 \tb'_0 - \rb^{-1}_a\right). \label{idR}
\end{equation}
The reflection and transmission matrices for each of the subsystems satisfy (\ref{TRmTR}). This allows us to show that
\begin{equation}
{\rb}{}'^{-1}_0(-k_3) \left(1 - {\rb}{}'_0 (-k_3){\rb}{}_a(-k_3)\right) {\tb}{}^{-1}_a (-k_3)\Rb'(-k_3){\rb}{}'^{-1}_a (-k_3) {\tb}_a (-k_3)= \rb'_0 (k_3)- \tb_0 (k_3)\rb^{-1}_0 (k_3)\tb'_0 (k_3)- \rb^{-1}_a(k_3)\,.\label{idRp}
\end{equation}
By using (\ref{SRR}), (\ref{idR}) and (\ref{idRp}) together with the identity $\det(1-AB)=\det(1-BA)$ we obtain
\begin{equation}
\det \S (k_3) = \frac{\det \left(1 - \rb'_0(-k_3) \rb_a(-k_3)\right)}{\det \left(1 - \rb'_0(k_3) \rb_a(k_3)\right)} \frac{\det \left(\rb_0 (k_3)\rb_a(k_3)\right)}{\det \left(\rb'_0 (-k_3)\rb'_a(-k_3)\right)}. \label{detSk3}
\end{equation}

We are interested in the dependence of Casimir energy on the distance $a$ only (so that we may neglect all $a$-independent parts). To analyze the $a$-dependence we should change a bit our notations. Let all reflection matrices related to the left subsystem carry a subscript $I$, while the matrices for the right subsystem from now on carry a subscript $\II$. Moreover, let us now move the boundary of second subsystem from $z=a$ to $z=0$. As a result, the reflection matrices receive a phase factor, and in new notation they read
\begin{equation}
\rb_{a}(k_3)=e^{2\ii ak_3}\rb_{\II}(k_3)\,,\qquad \rb_{a}'(k_3)=e^{-2\ii ak_3}\rb'_{\II}(k_3)\,.
\label{shiftr}
\end{equation}
Similarly, instead of $\rb_0$ ($\rb'_0$) we shall use $\rb_{I}$ ($\rb'_I$) in what follows.

The second fraction of \Ref{detSk3} reads in the new notation
\begin{equation*}
  \frac{\det \left(\rb_0 (k_3)\rb_a(k_3)\right)}{\det \left(\rb'_0 (-k_3)\rb'_a(-k_3)\right)}
  =\frac{\det \left(\rb_{I} (k_3)\rb_{\II}(k_3)\right)}{\det \left(\rb'_{I} (-k_3)\rb'_{\II}(-k_3)\right)}
\end{equation*}
showing no dependence on $a$. For computation of the distance dependent part of Casimir energy one may use
\begin{equation}
\ln \det \S =\ln \frac{\det \left(1 - e^{-2\ii ak_3}\rb'_{I}(-k_3) \rb_{\II}(-k_3)\right)}{\det \left(1 - e^{2\ii ak_3}\rb'_{I}(k_3) \rb_{\II}(k_3)\right)}\,.\label{lndetS}
\end{equation}

Let us change the integration variable in (\ref{ii_Lif1}) $k_3\to\omega=\sqrt{k_3^2+k_1^2+k_2^2}$, so that $d\omega=(k_3/\omega)dk_3$. Thus,
\begin{equation}
\mathcal{E}=- \frac1{4 \pi \ii}  \int \frac{d^2 k}{(2\pi)^2}\int_{\omega_0}^\infty d\omega\, \ln
\frac{\det \left(1 - e^{-2\ii ak_3}\rb'_{I}(-k_3) \rb_{\II}(-k_3)\right)}{\det \left(1 - e^{2\ii ak_3}\rb'_{I}(k_3) \rb_{\II}(k_3)\right)}\,,\label{Lif2}
\end{equation}
where $\omega_0=\sqrt{k_1^2+k_2^2}$. 

The momentum $k_3=\sqrt{\omega^2-k_1^2-k_2^2}$ in (\ref{Lif2}) becomes a dependent variable. The sign of phase of $k_3$ in the complex plain coincides with the sign of phase of $\omega$. This sign defines the regions where the exponents $e^{\pm 2iak_3}$ become damping, and thus it defines the direction in which the integration contour may be deformed. Hence,
\begin{eqnarray}
&&\mathcal{E}=- \frac1{4 \pi \ii}  \int \frac{d^2 k}{(2\pi)^2} \left[ \oint_{\gamma_-}d\omega
\ln \det \left(1 - e^{-2\ii ak_3}\rb'_{I}(-k_3) \rb_{\II}(-k_3)\right) \right. \nonumber\\
&&\qquad\qquad\qquad \left. - \oint_{\gamma_+}d\omega \ln \det \left(1 - e^{2\ii ak_3}\rb'_{I}(k_3) \rb_{\II}(k_3)\right) \right]\,. \label{Lif3}
\end{eqnarray}
The contour $\gamma_+$ starts at $\omega=\omega_0$, goes to $\omega=0$ above the real axis, and continues to $+\ii\infty$. $\gamma_-$ is just the reflection of $\gamma_+$ with respect to the real axis (Fig.\ \ref{fig:2n}(b)).

Since we agreed to neglect the contributions of discrete spectrum of $k_3$, the integrand has no poles between $0$ and $\omega_0$, so that the contributions from the pieces of $\gamma_+$ and $\gamma_-$ between these two points cancel against each other. If there were such discrete modes, their contributions should have been added as a finite sum to the $k_3$ integral in (\ref{ii_Lif1}). Then, such finite sums cancel the contribution of the poles, see \cite{Bordag1995,Bordag:2001qi,Nesterenko:2011ab}. The remaining parts of the integrals in (\ref{Lif3}) can be combined into a single integral
over imaginary frequencies, $\omega=\ii\xi$,
\begin{equation}
\mathcal{E}=\frac 1{4\pi} \int \frac{d^2 k}{(2\pi)^2}\int_{-\infty}^{\infty}d\xi\, \ln \det 
\left(1 - e^{-2ak_E}\rb'_{I}(\ii k_E) \rb_{\II}(\ii k_E)\right)\,.\label{Lif4}
\end{equation}
Here, $k_E\equiv \sqrt{\xi^2+k_1^2+k_2^2}$ and $k_3=\ii k_E$ for $\xi>0$, while $k_3=-\ii k_E$ for $\xi<0$. This is the Lifshitz formula for Casimir interaction of two planar systems we have been looking for. It has been derived under much weaker assumptions on the properties of the systems than in the previous literature. We stress the necessity to use the $\rb'$ matrix.  

\subsection{Transition to TE and TM modes}\label{sec:TETM}
The computations above have been done in the linear polarization basis. We used very little of peculiar properties of this basis, so that the Lifshitz formula remains valid in other polarizations as well. However, it is instructive to present explicit formulas for the transition to TE and TM modes. For large negative values of $z$ the components of electric field for TE and TM polarizations read
\begin{equation*}
\E^- =\frac{k_0}{k_1^2+k_2^2} \left\{ e^{\ii zk_3} \begin{pmatrix} k_2 \\ -k_1 \end{pmatrix} G^-_{\mathrm{TE},i} + e^{-\ii zk_3} \begin{pmatrix} k_2 \\ -k_1 \end{pmatrix} G^-_{\mathrm{TE},o} \right\}
+\frac{k_3}{k_1^2+k_2^2}\left\{ e^{\ii zk_3} \begin{pmatrix} -k_1 \\ -k_2 \end{pmatrix} G^-_{\mathrm{TM},i} + e^{-\ii zk_3} \begin{pmatrix} k_1 \\ k_2 \end{pmatrix} G^-_{\mathrm{TM},o} \right\}
\end{equation*} 
where $G_{\mathrm{TE (TM)}}$ are coefficients of expansion in TE,TM basis. 
This yields
\begin{eqnarray}
&&\E^-_i = \frac{k_0}{k_1^2+k_2^2} \begin{pmatrix} k_2 \\ -k_1 \end{pmatrix} G^-_{\mathrm{TE},i}
+ \frac{k_3}{k_1^2+k_2^2} \begin{pmatrix} -k_1 \\ -k_2 \end{pmatrix} G^-_{\mathrm{TM},i},\label{EmEMi}\\
&&\E^-_o = \frac{k_0}{k_1^2+k_2^2} \begin{pmatrix} k_2 \\ -k_1 \end{pmatrix} G^-_{\mathrm{TE},o}
+ \frac{k_3}{k_1^2+k_2^2} \begin{pmatrix} k_1 \\ k_2 \end{pmatrix} G^-_{\mathrm{TM},o}.\label{EmEMo}
\end{eqnarray}
To obtain the corresponding formulas at large positive $z$, it is enough to replace the superscript $+$ with $-$ and interchange the subscripts $i$ and $o$ in (\ref{EmEMi}) and (\ref{EmEMo}). Thus we conclude that
\begin{eqnarray}
&& (\E_o^-,\E^+_o)^T=U_o(G^-_{\mathrm{TE},o},G^-_{\mathrm{TM},o},G^+_{\mathrm{TE},o},G^+_{\mathrm{TM},o})^T,\label{EG1}\\
&& (\E_i^-,\E^+_i)^T=U_i(G^-_{\mathrm{TE},i},G^-_{\mathrm{TM},i},G^+_{\mathrm{TE},i},G^+_{\mathrm{TM},i})^T,\label{EG2}
\end{eqnarray}
with
\begin{equation}
U_i=\begin{pmatrix} \wb (k_3) & 0 \\ 0 & \wb(-k_3) \end{pmatrix}\,,\qquad
U_o=\begin{pmatrix} \wb (-k_3) & 0 \\ 0 & \wb(k_3) \end{pmatrix}\,,\qquad
\wb=\frac 1{k_1^2+k_2^2} \begin{pmatrix} k_0 k_2 & -k_3k_1 \\ -k_0k_1 & -k_3k_2 \end{pmatrix}.
\end{equation}

By using these formulas, one may express the scattering matrix in TE-TM polarizations, $\S_{\EM}$, through the scattering matrix in linear polarizations, $\S_{\XY}$, as
\begin{equation}
\S_{\EM}=U_o^{-1} \S_{\XY} U_i\,.\label{SUSU}
\end{equation}
In particular,
\begin{equation}
\Rb_{\EM}=\wb^{-1}(-k_3) \Rb_{\XY} \wb(k_3),\qquad \Rb_{\EM}'=\wb^{-1}(k_3) \Rb_{\XY}' \wb(-k_3).\label{REMRXY}
\end{equation}
Of course, the same relations hold for the reflections matrices $\rb_{I}$ and $\rb_{\II}$ of the subsystems. The Lifshitz formula (\ref{Lif4}) does not change under the change of the basis, as expected.

\subsection{Dielectric substrates}\label{sec:diel}

Above in this section we considered the Casimir interaction of planar systems surrounded by the vacuum. This, of course, includes finite thickness dielectric slabs. However, in many cases it is useful to consider half-infinite dielectric substrates. Here we briefly describe the modification in our procedure required by this case.

Note, that the formulas derived so far can be used for a rather general permittivity tensor (as, e.g., for the photonic TIs, see \cite{Fuchs:2017uyg}). From now on we restrict ourselves to scalar permittivities and permeabilities proportional to a unit matrix.

First of all, let us consider what happens with the Lifshitz formula (\ref{Lif4}) in the infinite thickness limit. Let us take one of the subsystems, say $I$, and consider it as a composite system consisting of two interfaces at $z=0$ and $z=-\ell$ separated by a dielectric. Then, $\rb'_I$ can be expressed through the scattering matrices at the interfaces by using the formula on the line above (\ref{idR}) (with obvious modifications). The dependence on $\ell$ is given by analogs of Eqs.\ (\ref{shiftr}). Then, after the Wick rotation, one gets (\ref{Lif4}) with $\rb'_I$ being the reflection matrix at the interface at $z=0$ between the vacuum and the dielectric. The corrections are exponentially suppressed at $\ell\to\infty$. A similar procedure can be performed with $\rb_{\II}$. The corresponding scattering matrices are computed in Sec.\ \ref{sec:refl}.

Secondly,  the transformation \Ref{REMRXY} of the reflection coefficients $\rb$ to the TE-TM basis has to be modified. We note that the  $\rb_I$ describes the reflection process in the leftmost medium, where the dielectric permittivity and magnetic permeability are denoted by $\epsilon^-$, $\mu^-$.  $\rb_{\II}'$ stands for reflection into the rightmost medium with $\epsilon^+$, $\mu^+$, respectively. It is easy to see that the expansion of the electric field in TE, TM modes in medium is obtained from (\ref{EmEMi},\ref{EmEMo}) by substituting $k_3$ by corresponding wave vector in the medium. 
Thus, instead of $\wb(k_3)$ in \Ref{REMRXY} one has use $\wb(k_3^-)$ and $\wb(k_3^+)$  for  $\rb_{I}$ and $\rb_{\II}'$ correspondingly, with 
$$
	k_3^\pm=\sqrt{ \epsilon^\pm\mu^\pm k_0^2-k_1^2-k_2^2}.
$$
For $\rb_{I}'$ and $\rb_{\II}$ which both describe the reflection into the vacuum  between two dielectrics we shall continue to use  $\wb(k_3)$. Summarizing, the transformation rule reads,
\begin{eqnarray}
& \rb_{I, \EM}=\wb^{-1}(-k_3^-)\, \rb_{I, \XY} \, \wb(k_3^-),\qquad
   & \rb_{I, \EM}'=\wb^{-1}(k_3) \, \rb_{I, \XY}'\,  \wb(-k_3),
   \label{rEMrXY-diel-1}
  \\
& \rb_{\II, \EM}=\wb^{-1}(-k_3) \, \rb_{\II, \XY}\,  \wb(k_3),\qquad
  & \rb_{\II, \EM}'=\wb^{-1}(k_3^+) \, \rb_{\II, \XY}'\,  \wb(-k_3^+).
  \label{rEMrXY-diel-2}
\end{eqnarray}
Under these transformations, the Lifshitz formula remains invariant.

\section{Casimir interaction of surfaces with P-odd conductivity}\label{sec:p-odd}

\subsection{Reflection matrix}\label{sec:refl}
\begin{figure}
	\includegraphics[width=5.8truecm]{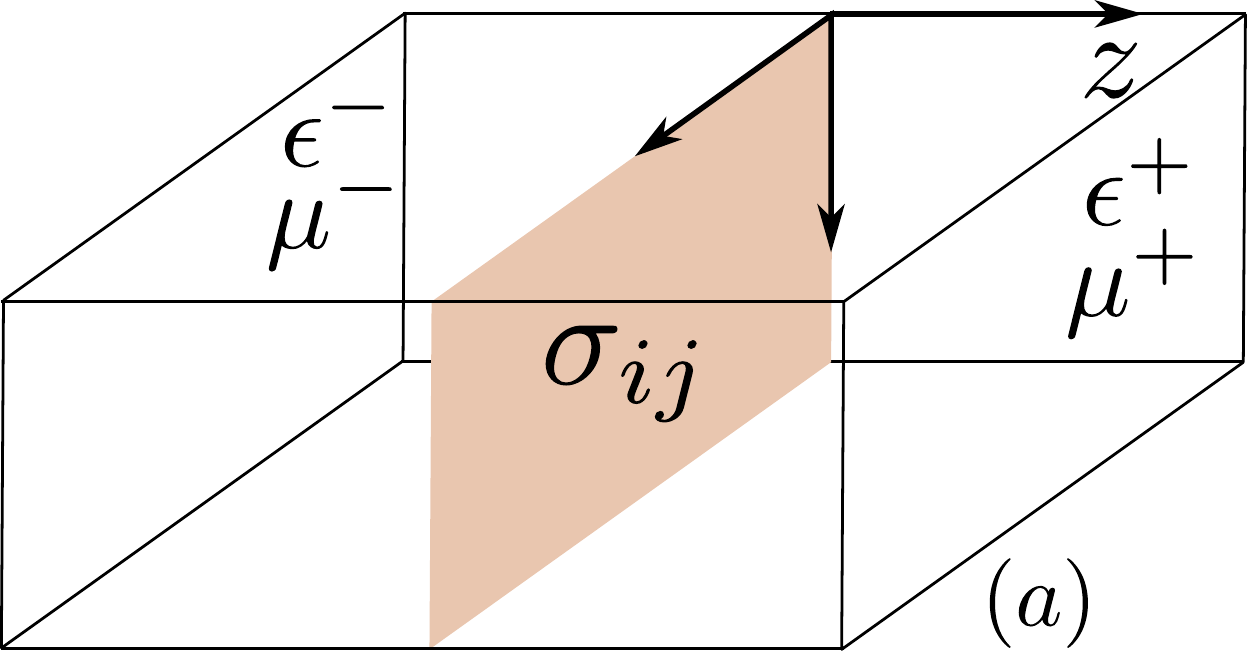}\hspace{6em}\includegraphics[width=8truecm]{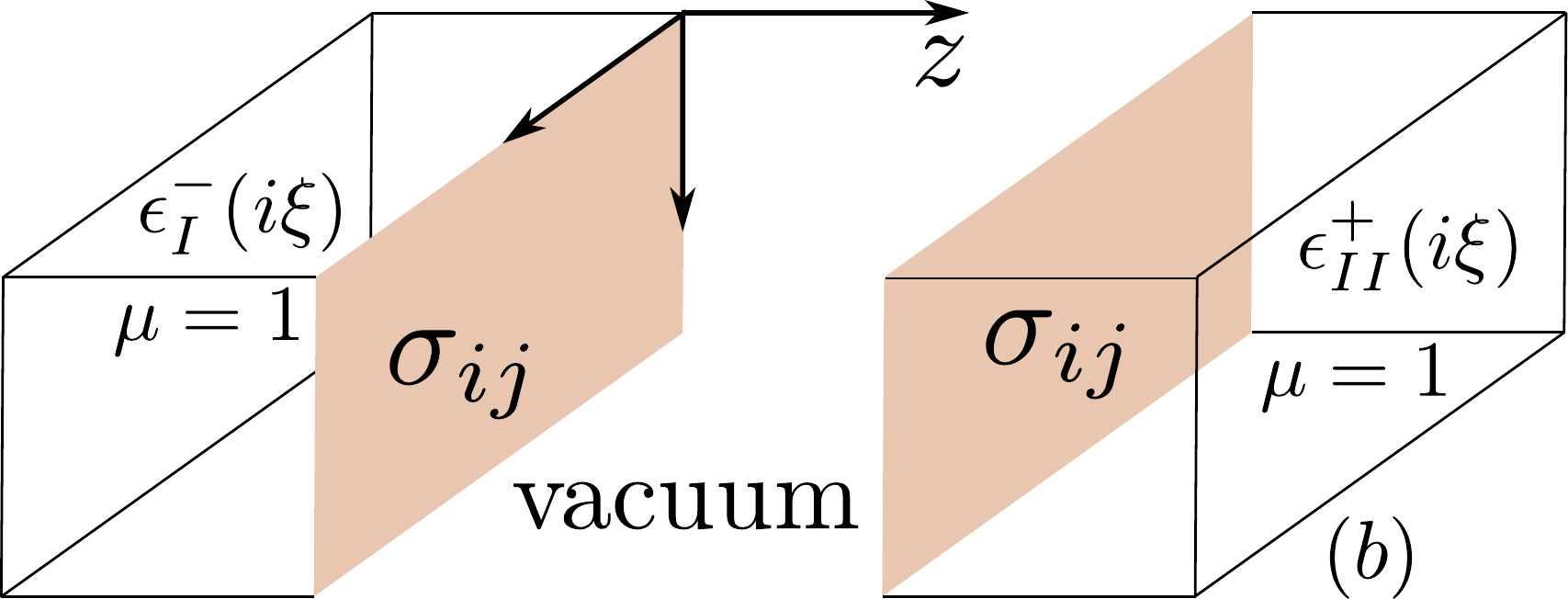}
	\caption{The configurations used in Sections \ref{sec:refl} and \ref{sub:ano}: two dielectrics separated by a plane conducting surface (panel (a)) and two dielectrics with conducting boundaries separated by a vacuum gap (panel (b)).}\label{fig:3n}
\end{figure}

To describe the interaction of a planar TI  with an electromagnetic field we consider a plane characterized by a surface conductivity tensor $\si_{ij}$, $i,j=1,2$, sandwiched between two materials with dielectric permittivities $\epsilon^\pm$ and magnetic permeabilities $\mu^\pm$, Fig.\ \ref{fig:3n}(a). If we assume that the dielectric substrates extend to infinity at both sides of the conducting surface, we can construct the reflection coefficient following, for instance, the techniques of \cite{Bordag:2017vrt}, in relatively compact and transparent form. The latter paper dealt with isolated surface with symmetric conductivity, while in the case of TIs we should not restrict ourselves in this respect. 

Summarizing the approach, one is to solve the scattering problem for electromagnetic field subject to the following matching conditions at the interface $z=0$
\begin{eqnarray}
&&[E_a]=[B_3]=0\,, \quad a=1,2\label{mEa}\\
&&[D_3]=\frac{k_1 E_1 \si_{11}
		+(k_2 E_1 +k_1 E_2)\si_{12}
			+ k_2 E_2 \si_{22}}{k_0},\label{mE3}\\
&&[H_1]=\si_{12}E_1+\si_{22}E_2,\label{mH1}\\
&&[H_2]=-\si_{11}E_1-\si_{12}E_2,\label{mH2}
\end{eqnarray}
where $\D=\epsilon \E$, ${\bm H} = \mu^ {-1} \B$, and the square brackets denote the discontinuity of corresponding function on the interface. The scattering electric field is defined as
\be
	\E=e^{\ii k_a x^ a}\left\{
		\begin{array}{ll}
		\E^-_i e^{\ii k_3^ - z}
			+\E^-_o e^{-\ii k_3^ - z}, &z<0\\
		\E^+_i e^{-\ii k_3^+ z}
			+\E^+_o e^{\ii k_3^+ z}, &z>0 ,
		\end{array}
		\right.
\ee
while the magnetic components are derived through the Maxwell equations,
\be
  \B=\frac{-\ii}{k_0} \nabla \times \E ,\qquad 
    \nabla \equiv (\partial_1,\partial_2,\partial_3).
\ee
After long but straightforward calculations we arrive at the following result for the scattering matrix \Ref{Smat2} in XY basis
\begin{equation}
	\rb_\XY= \frac 1{\Delta }
	\(\begin{array}{cc}
		\rho_{11} & \rho_{12}\\
		\rho_{21} & \rho_{22}
		\end{array}
		\),
		\label{R-expl}
\end{equation}
\begin{eqnarray}
\rho_{11}&=&\frac{\ep^-}{\mu^-}-\frac{\ep^+}{\mu^+}
	+\sigma_{12}\si_{21}
	-\sigma_{11}\sigma_{22} 
	+\frac{k_1k_2}{k_0}\(\frac{\si_{12}+\si_{21}}{\mu^+ k_3^+}-\frac{\si_{12}-\si_{21}}{\mu^- k_3^-}\)\nonumber\\
&&	+\frac{\si_{22}}{k_0}\(\frac{(k_3^-)^2+k_1^2}{\mu^- k_3^-}-\frac{(k_3^+)^2+k_1^2}{\mu^+ k_3^+}\)
	-\frac{\si_{11}}{k_0}\(\frac{(k_3^-)^2+k_2^2}{\mu^- k_3^-}+\frac{(k_3^+)^2+k_2^2}{\mu^+ k_3^+}\)\nonumber\\
&&	-\frac{\ep^+\mu^+-\ep^-\mu^-}{\mu^+\mu^-}\frac{k_2^2-k_1^2}{k_3^-k_3^+}
\,,\nonumber\\
\rho_{12}&=& 2
	\( 
	\frac{\ep^+\mu^+-\ep^-\mu^-}{\mu^+\mu^-}\frac{k_1 k_2}{k_3^-k_3^+}
	+ \frac{k_1k_2}{k_0} \frac{\si_{22}}{\mu^- k_3^-}
	- \frac{\si_{12}}{k_0} \frac{(k_3^-)^2+k_2^2}{\mu^- k_3^-}
	\),
\nonumber\\
\rho_{22}&=& \rho_{11}(1\leftrightarrow2) ,\qquad  \rho_{21}=\rho_{12}(1\leftrightarrow2), 
\nonumber
\end{eqnarray}
and
\begin{eqnarray}
\Delta &=& \frac{\ep^-}{\mu^-}+\frac{\ep^+}{\mu^+}
    + \frac{ \ep^+ k_3^-}{\mu^- k_3^+}+ \frac{ \ep^- k_3^+}{\mu^+ k_3^-}
    -\sigma_{12}\si_{21}
	+\sigma_{11}\sigma_{22} 
    +\frac{\si_{11}}{k_0}\(\frac{(k_3^-)^2+k_2^2}{\mu^- k_3^-}+\frac{(k_3^+)^2+k_2^2}{\mu^+ k_3^+}\)
    \\
    &&
    +\frac{\si_{22}}{k_0}\(\frac{(k_3^-)^2+k_1^2}{\mu^- k_3^-}+\frac{(k_3^+)^2+k_1^2}{\mu^+ k_3^+}\)
    -\frac{k_1k_2}{k_0}(\si_{12}+\si_{21})\(\frac{1}{\mu^+ k_3^+}+\frac{1}{\mu^- k_3^-}\),
    \nonumber
\end{eqnarray}
where $k_3^\pm=\sqrt{ \epsilon^\pm\mu^\pm k_0^2-k_1^2-k_2^2}$ . A similar, though a bit less general expressions for the reflection amplitudes have been derived in \cite{Tse:2011}.

Particularly interesting is that in this basis
\be
  \rb_\XY'= \rb_\XY,\qquad 
  \tb_\XY=1+\rb_\XY,
  \label{rXY'}
\ee
while neither equality holds in the TE-TM basis. Applying (\ref{rEMrXY-diel-1}) and (\ref{rEMrXY-diel-2}) to \Ref{R-expl} we deduce that 
\be
  \rb_\EM'(\ep^-, \mu^-,\ep^+, \mu^+;\sigma)
    = \rb_\EM(\ep^+, \mu^+,\ep^-, \mu^-;\sigma^T)
\ee
if the following condition on conductivities is satisfied:
\be
	(k_1^2-k_2^2)(\si_{12}+\si_{21})-2 k_1 k_2 (\si_{11}-\si_{22}) =0.\label{invEM}
\ee

\subsection{Particular case: constant conductivities}\label{sec:part}

Let us consider scattering on two planes in the vacuum, $\epsilon^\pm=\mu^\pm=0$. In this case, the matrix $\rb_\XY$ can be represented in the following compact form:
\begin{equation}
\rb_\XY = -\frac{2k_0^2 \sigmab - 2\kb \otimes \kb\sigmab + {\bf 1}_2 k_3k_0 \det\sigmab}{2k_0^2 \tr \sigmab - 2\kb\kb\sigmab + k_3k_0 (4 + \det\sigmab) }, 
\end{equation}
where $\bm{k} = (k_1,k_2)$.  Taking into account simple algebra for arbitrary $2\times 2$ matrices $\mathbf{A}$ and $\mathbf{A}'$ and a scalar $\nu$
\begin{gather*}
\det \left(\bm{1}_2 - \nu \mathbf{A} \mathbf{A}'\right) = 1 - \nu \tr (\mathbf{A} \mathbf{A}') + \nu^2 \det (\mathbf{A} \mathbf{A}'),\\
\tr (\mathbf{A} \mathbf{A}') = \tr (\mathbf{A}) \tr(\mathbf{A}') + \det \mathbf{A} + \det \mathbf{A}' - \det (\mathbf{A} + \mathbf{A}'),
\end{gather*}
and Eq. \eqref{Lif4} we obtain the following expression for the Casimir energy
\begin{gather}
\mathcal{E} = \int \frac{d^2 k}{(2\pi)^3} \int_0^\infty d\xi \ln \left(1 - e^{-2 a k_E}\left(\frac{\xi^2 k_E^2 }{bb'} \left[(4- \det\sigmab) (4- \det\sigmab' ) + 4 \det(\sigmab - \sigmab') \right] - \frac{4\xi k_E}{b}   - \frac{4\xi k_E}{b'}   + 1 \right)\right. \nonumber \\
+\left.  e^{-4 a k_E} \frac{\xi^2 k_E^2 }{bb'} \det\sigmab \det \sigmab'\right),\label{eq:XY}
\end{gather}
where the integration variable $\xi$ play the role of an imaginary frequency, $k_E\equiv \sqrt{\xi^2+\kb^2}$ is the total momentum parallel to the surface Wick-rotated to the Euclidean region, and $b = 2\xi^2 \tr \sigmab + 2 \kb\kb\sigmab  + \xi k_E \bigl(4+ \det\sigmab\bigr),\ b' =  2\xi^2 \tr \sigmab' + 2 \kb\kb\sigmab'  + \xi k_E \bigl(4+ \det\sigmab'\bigr)$. 

The expression above holds for arbitrary tensors of conductivity.  Let us consider a particular simple case of constant conductivities on both plates with the $\sigma_{ij} = 2(\nxi \delta_{ij} + \eta \epsilon_{ij})$,  where the positive constants $\nxi$ and $\eta$ define isotropic dc longitudinal and Hall conductivities, respectively. The constant conductivity approximation can be used at large separations $a$. The critical value of $a$ crucially depends on the model of conductivity. For example, in framework of the Drude-Lorentz model of graphene's conductivity \cite{Khusnutdinov:2015:cefasocp} the Drude-Lorentz nature of the response becomes less relevant for $a\gg 14nm$.  Our constant $\eta$ coincides with the Chern-Simons  parameter $a$ used in the papers \cite{Markov:2006js,Marachevsky:2017tdo}. For the second plane,  we consider the following three possibilities: i) a ``symmetric" plane with the same conductivity tensor $\sigmab' = \sigmab$, and ii) an ``antisymmetric" plane with the opposite sign of Hall conductivity $\sigma'_{ij} = 2(\nxi \delta_{ij} - \eta \epsilon_{ij})$, and iii) an ideal conductor with $\sigma'_{ij} = 2\nxi' \delta_{ij}$ in the limit $\nxi'\to \infty$. We shall mark these cases as ``$s$''\/, ``$a$''\/ and ``$id$'', respectively.  

Taking into account the general expression \eqref{eq:XY}, we obtain the Casimir energy 
\begin{equation}
\mathcal{E}_{n} = -\frac{1}{16\pi^2a^3} \int_0^1 \left\{\Li_4 (Z^+_{n})+\Li_4 (Z^-_{n})\right\} dx,
\end{equation}
where $n=s,a,id$, and $\Li_4(Z)$ is a polylogarithm function with the following arguments:
\begin{eqnarray}
 Z^\pm_s &=& \frac{1}{2}\left(\nxi ^2 (x^2+1)^2 + 2 x (\eta^2+\nxi^2) (\nxi  (x^2+1)+x (\eta^2+\nxi ^2-1))\pm \sqrt{\nxi^2 (1-x^2)^2-4 \eta ^2 x^2} (\nxi  (x^2+1)+ 2 x (\eta ^2+\nxi ^2))\right)\nonumber \\
   &\times& \left(\nxi  \left(x^2+1\right)+x \left(\eta ^2+\nxi ^2+1\right)\right)^{-2},\nonumber\\
Z^\pm_a &=& \frac{1}{2}\left(x^2 ((\eta ^2+\nxi ^2)^2 + 2 (\eta ^2-\nxi ^2)) + (\nxi  (x^2 + 1) + x (\eta^2 + \nxi^2 ))^2 \pm \nxi  (1-x^2) \sqrt{(\nxi  (x^2+1)+2 x (\eta ^2 + \nxi^2))^2+4 \eta ^2 x^2} \right) \nonumber\\
 &\times& \left(\nxi  \left(x^2+1\right)+x \left(\eta ^2+\nxi ^2+1\right)\right)^{-2},\nonumber\\
Z^\pm_{id} &=& \frac{\nxi  (x^2+1)+2 x (\eta ^2+\nxi ^2) \pm \sqrt{\nxi ^2 (1 - x^2)^2-4 \eta ^2 x^2}}{2 (\nxi  (x^2+1)+x (\eta ^2+\nxi ^2+1))}.
\end{eqnarray}

In the case of pure Hall conductivity, $\nxi =0$, we recover the Casimir energy found in Refs. \cite{Markov:2006js,Marachevsky:2017tdo}
\begin{equation}
 \mathcal{E}_s = -\frac{1}{8\pi^2a^3} \Re \Li_4 \left(\frac{\eta^2}{(\eta + \ii)^2}\right),\ \mathcal{E}_a = -\frac{1}{8\pi^2a^3}  \Li_4 \left(\frac{\eta^2}{\eta^2 + 1}\right),
\end{equation}
and for the case without Hall conductivity, $\eta=0$, we obtain
\begin{equation}
 \mathcal{E}_{s,a} = -\frac{1}{16\pi^2a^3} \int_0^1 \left\{\Li_4 \left(\frac{\nxi^2}{(x + \nxi)^2}\right) + \Li_4 \left(\frac{x^2 \nxi^2}{(1 + x \nxi)^2}\right)\right\}dx ,
\end{equation}
in agreement with Ref. \cite{Khusnutdinov:2014:Cefswcc,*Khusnutdinov:2015:cefasocp,*Khusnutdinov:2015:Cefacopcs,*Khusnutdinov:2016:cpefasocp,*Khusnutdinov:2018:tcacpii$n$p2dm}. For ideal case we have
\begin{equation}
 \mathcal{E}_{id}(\nxi=0) = -\frac{1}{8\pi^2a^3} \Re \Li_4 \left(\frac{\eta}{\eta + \ii}\right),\ \mathcal{E}_{id}(\eta =0) = -\frac{1}{16\pi^2a^3} \int_0^1 \left\{\Li_4 \left(\frac{\nxi}{x + \nxi}\right) + \Li_4 \left(\frac{x \nxi}{1 + x \nxi}\right)\right\}dx.
\end{equation}

\begin{figure}
	\includegraphics[width=6cm]{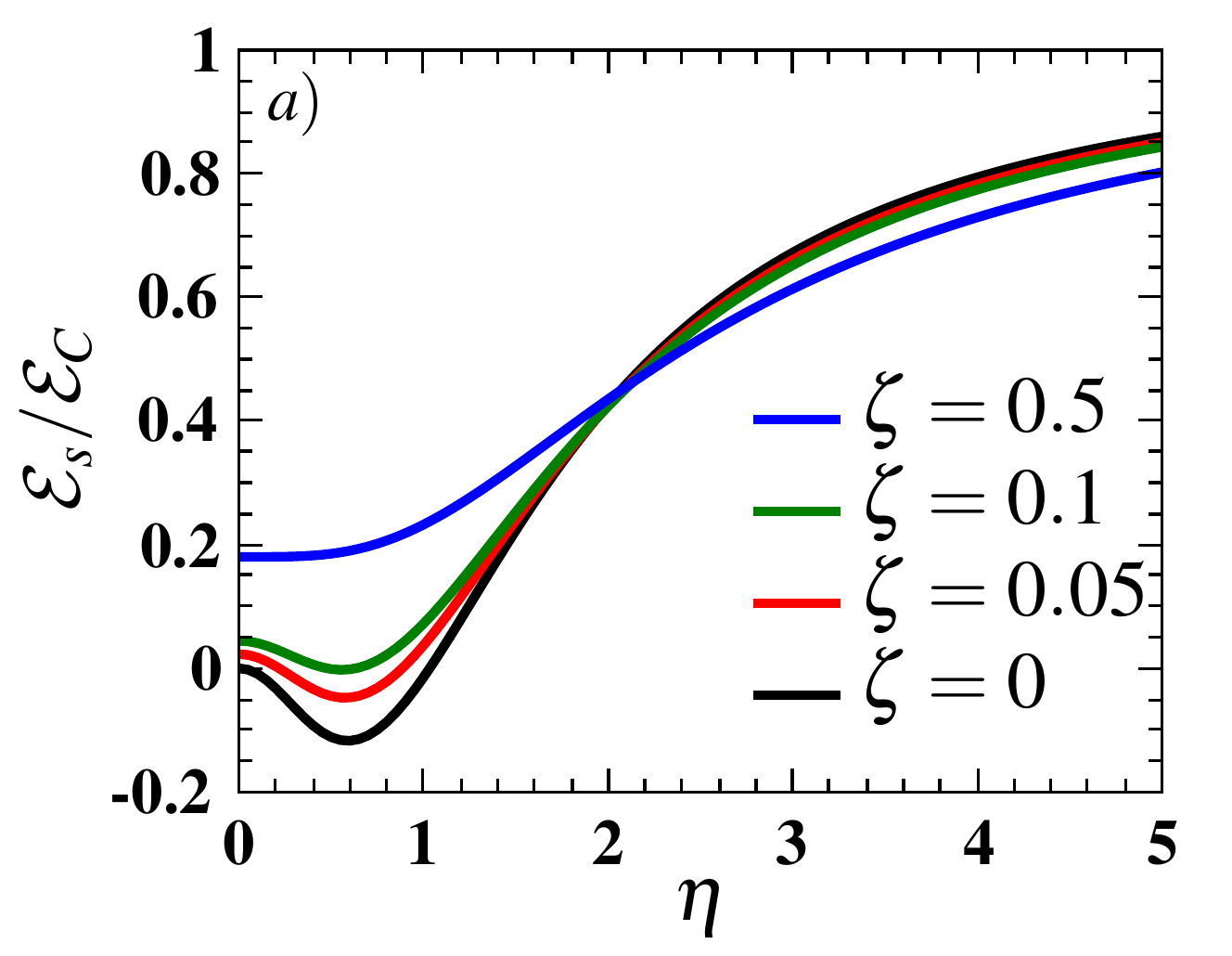}\includegraphics[width=6cm]{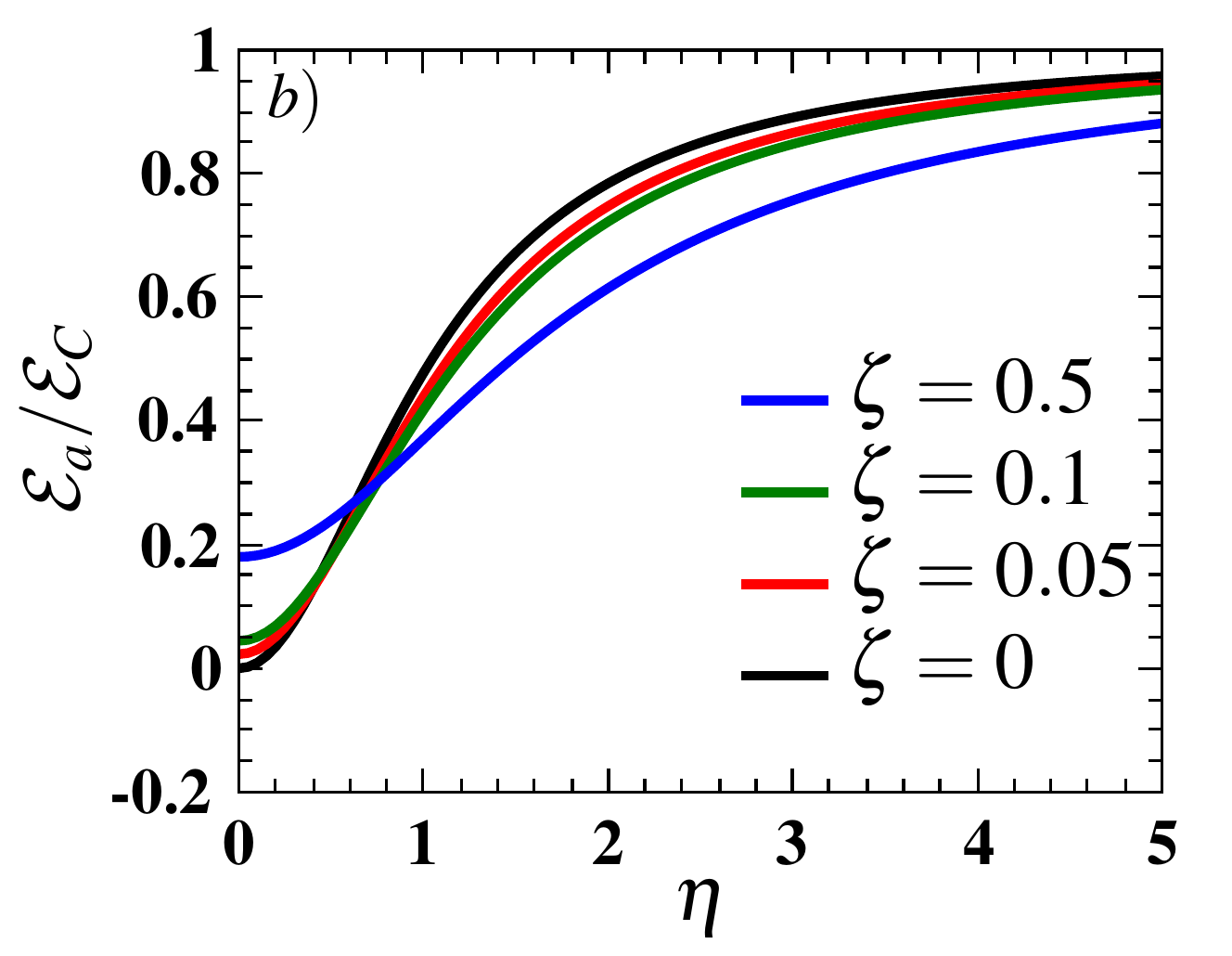}\includegraphics[width=6cm]{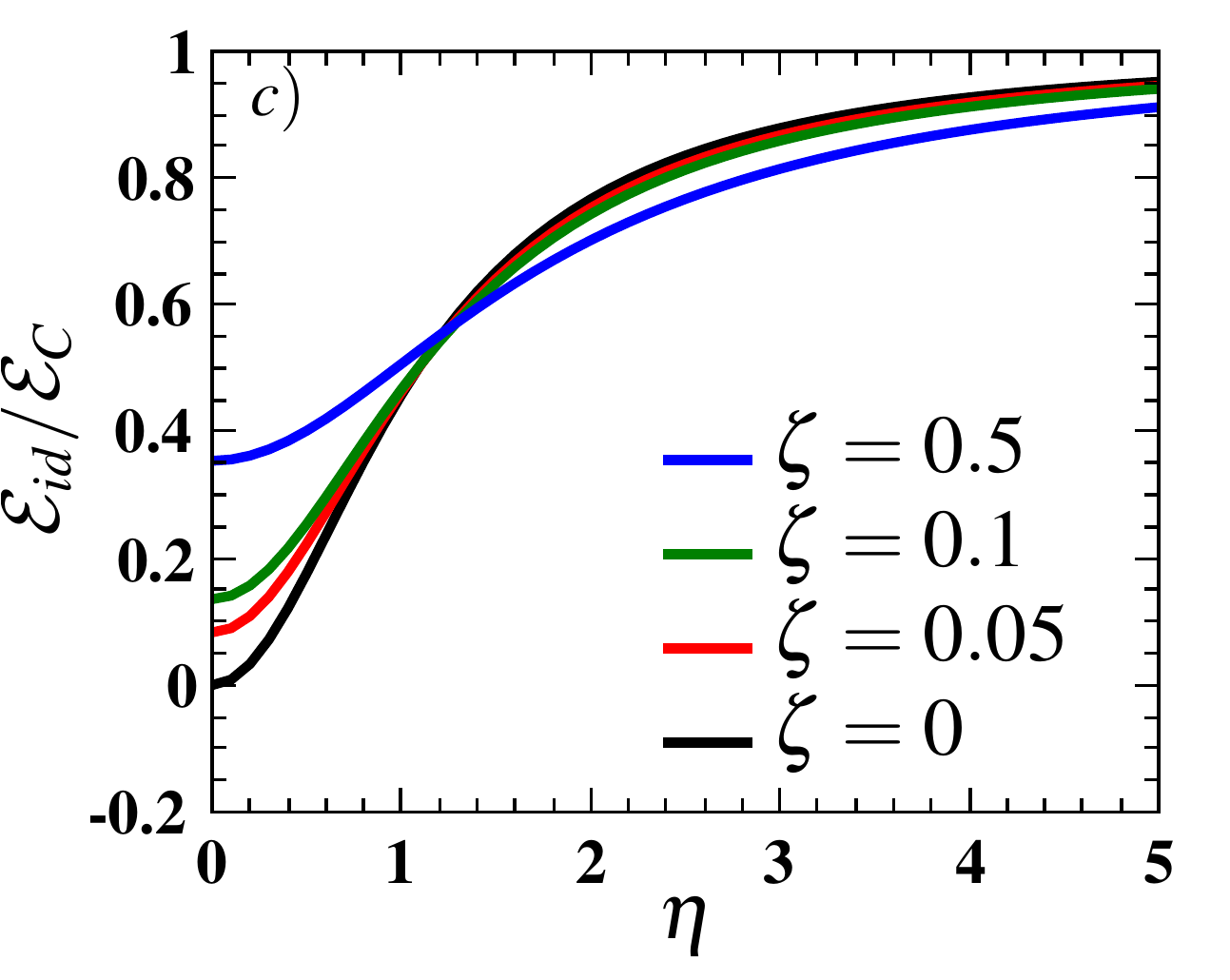}
	\caption{The Casimir energy $\mathcal{E}_{s,a, id}$ normalized to the Casimir energy  for two ideal conducting planes $\mathcal{E}_C = -\frac{\pi^2}{720a^3}$. The energy for symmetric case with two identical planes has negative value (repulsive force) for dc conductivity $\nxi < 0.103$ (panel a). The Casimir energy for two planes with opposite Hall conductivity is always positive which produces attractive force (panel b). The same behavior is observed for case if one plane is taken to be an ideal conductor (panel c).}\label{fig:1}
\end{figure}

The sign of energy in the symmetric case $\mathcal{E}_s$ crucially depends on the value of longitudinal conductivity $\nxi$. In the panel a) of Fig. \ref{fig:1} we show dimensionless Casimir energy $\mathcal{E}_s/\mathcal{E}_C$ as function of Hall conductivity $\eta$ for different values of longitudinal conductivity $\nxi$. Here $\mathcal{E}_C = -\frac{\pi^2}{720a^3}$ is the Casimir energy for two ideal planes. The higher value of $\nxi$ the narrower domain of Hall conductivity $\eta$ where energy $\mathcal{E}_s/\mathcal{E}_C$ is negative  (repulsive force) and this domain  disappears for $\nxi > 0.103$. The Casimir dimensionless energy $\mathcal{E}_a/\mathcal{E}_C$ for antisymmetric case (panel b) and $\mathcal{E}_{id}/\mathcal{E}_C$ for ideal case  (panel c) are always positive and produce attractive force for any value of Hall conductivity. 
 
Regarding the panel c) of Fig.\ \ref{fig:1} we note that the dependence of Casimir force of Hall conductivity parameter $\eta$ is same strong as for the force between two Hall surfaces. The case when one of the surfaces is a good conductor is much easier to realize at an experiment, see \cite{Bordag:2009zzd}.

To parametrize the  surface conductivity, papers \cite{Grushin:2011b,Grushin:2011,RodriguezLopez:2011dq} used angle $\theta$ in the topological action term $\sim \theta \int d^4x \E \cdot \B$ and obtained repulsion when the angles  in two interaction bodies were of opposite sign. Integration by parts in the $\theta$-term introduces a relative minus sign between Hall conductivities $\sigma_{12}$ of the interacting surfaces, depending on whether the surface corresponds to an upper or a lower limit of the integration over $z$. Thus, the Casimir repulsion in \cite{Grushin:2011b,Grushin:2011,RodriguezLopez:2011dq} corresponds to Hall conductivities of the same sign, exactly as in our computations above in this section. 

The situation with Chern insulators \cite{Rodriguez-Lopez:2013pza} is different. In the low frequency limit the Hall conductivity of Chern insulators is proportional to the Chern number without any additional sign factor (see Eq.\ (41) in the Supplementary material to Ref.\ \cite{Rodriguez-Lopez:2013pza}), while the longitudinal conductivity vanishes in this limit. Therefore, one may expect the large distance tail of the Casimir energy to be consistent with our analysis. Indeed, our Fig. \ref{fig:1}a,b for $\nxi=0$ is very similar to Fig.2b of \cite{Rodriguez-Lopez:2013pza}, but with relative signs of Hall conductivities flipped! We attribute this discrepancy to the use of the Lifshitz formula with $\Rb$ instead of $\Rb'$, see \cite[Eq. (1)]{Rodriguez-Lopez:2013pza}.

\subsection{The role of parity anomaly}\label{sub:ano}
The Chern-Simons term on the surface of topological insulators is usually associated with specific boundary states (see \cite{Tkachov2013} for a detailed discussion of the dynamics of such states). It seems natural therefore to describe surface conductivity of topological insulators through the polarization tensor of $2+1$ dimensional fermions, as in \cite{Chen:2011,Chen:2012ds}. This approach was applied previously to the Casimir interaction of graphene in, e.g., \cite{Bordag:2009fz,Fialkovsky:2011pu,Bordag:2015zda}. Moreover, just this approach has been confirmed experimentally \cite{Banishev:2013} according to the analysis of \cite{Klimchitskaya:2014axa}. Thus, the general framework of the papers \cite{Chen:2011,Chen:2012ds} looks very natural to us, but there are two points that we like to address here. One point is the use in \cite{Chen:2011,Chen:2012ds} of a version of the Lifshitz formula without $\rb'$. This is easy to correct as we have already explained above. The other point is more subtle. It is  related to the parity anomaly that was not included in the expressions for conductivity in \cite{Chen:2011,Chen:2012ds}.

The parity anomaly in fermionic quantum field theories in odd-dimensions was discovered mid 1980s \cite{Niemi:1983rq,Redlich:1983dv,AlvarezGaume:1984nf}, see \cite{Dunne:1998qy} for a review. Roughly speaking, the parity anomaly means that any quantization of $2+1$ dimensional fermions that is invariant under large gauge transformations inevitably breaks the parity invariance leading to a Chern-Simons term in the effective action. Whether these arguments fully apply to the edge states of topological insulators is still under a debate, see \cite{Mulligan:2013he,Zirnstein:2013tba,Konig:2014ema}.  A reliable way to sort out the problem is to perform quantum calculations on a $3+1$ dimensional space with a boundary. There is a single computation of this kind\footnote{The paper \cite{Mulligan:2013he} treats domain wall fermions rather that the boundary states. The former ones are allowed to propagate beyond the boundary and carry considerably more degrees of freedom than the latter ones.} \cite{Kurkov:2017cdz}. This paper considered massless fermions only. Even though the presence of the anomaly was confirmed in \cite{Kurkov:2017cdz}, the problem cannot be considered as completely settled so far. At any rate, there is a strong physical argument in favor of the anomaly: one expects that the contribution of boundary modes to conductivity vanishes when these modes become infinitely massive. This property requires the presence of anomaly term in the Hall part of conductivity.  

We conclude that there are good grounds to believe that the parity anomaly term is present in the conductivity generated by boundary modes. Below we study the influence of this term on Casimir interaction.

The conductivities generated by a single planar quasirelativistic fermion may be found in \cite{Fialkovsky:2011wh,*Fialkovsky:2012:qftig} (together with the references to original calculations). They are expressed through two scalar form factors $\Psi$, $\phi$  and read in the Minkowski signature
\begin{eqnarray}
&&  \si_{11} = \frac{i\al (k_0^ 2-v_F^2 k_2^2) \Psi(\tk)}{k_0 \tk^2}, \qquad
  \si_{22} = \frac{i\al (k_0^ 2-v_F^2 k_1^2) \Psi(\tk)}{k_0 \tk^2}, \nonumber\\
&&  \si_{12} =-\al \phi (\tk)+ \frac{i \al v_F^2 k_1 k_2 \Psi(\tk)}{k_0 \tk^2}, \qquad
  \si_{21} =+\al \phi (\tk) + \frac{i \al v_F^2 k_1 k_2 \Psi(\tk)}{k_0 \tk^2},
  \label{iii_sigmas}
\end{eqnarray}
where $\tk^2=k_0^2 - v_F^2(k_1^2+k_2^2)$, $v_F$ is the Fermi velocity and $\alpha=1/137$ is the fine structure constant.  After transition to the Euclidean signature, $\tk \to \tk_E =+\sqrt{\xi^2 +v_F^2(k_1^2+k_2^2)}$, the form factors are written as
\begin{eqnarray}
&&\Psi_E(\tk_E)=
        \frac{2 m \tk_E+ (\tk_E^2-4m^2){\rm arctan}({\tk_E}/{2m})}{2\tk_E},\qquad
    \label{Psi_E}\\
&&
\phi_E(\tk_E)
    =\frac{2m\, {\rm arctan}(\tk_E/2 m )}{\tk_E}-{\mathds A}
    \label{phi_E},
\end{eqnarray}
with $m$ being  mass of the fermion. 
The anomaly contribution is denoted by ${\mathds A}$.  The anomaly is present when  ${\mathds A}=1$. 

We stress, that the mass $m$ in (\ref{Psi_E}) and (\ref{phi_E}) was positive. For a negative mass, one has to change the overall sign of $\phi_E$, while $m$ has to be understood as an \emph{absolute value} of the mass. All in all, to obtain the conductivity generated by $n^+$ modes with the mass $+m$ and $n^-$ modes with $-m$ one has to replace
\begin{equation}
\Psi_E \to (n^++n^-)\Psi_E ,\qquad \phi_E \to (n^+-n^-)\phi_E\,.\label{npnm}
\end{equation}
For a single surface mode or for several modes of the same mass the sign of the mass corresponds to the sign of the Hall conductivity.

The authors \cite{Chen:2011,Chen:2012ds} who worked within a similar model with ${\mathds A}=0$ used a topological angle $\theta$ to discuss the Hall conductivity. Since $\phi_E$ depends on the momenta in a non-trivial way, the relation between topological action in the bulk and conductivity on the boundary is less clear than in the case of a constant Hall conductivity. They had to define their $\theta$ independently, and assumed the sign of $\theta$ being tantamount to the sign of Hall conductivity\footnote{We find this identification somewhat misleading since in the standard case the sign of $\theta$ cannot be immediately translated into the Hall conductivity. As we explained at the end of Section \ref{sec:part}, one should take into account a possible sign factor from the integration by parts. Quite possibly, this has led to a confusion in the comparison of the results of \cite{Chen:2011} to \cite{Grushin:2011b}.}, see the paragraph below Eq.\ (12) in \cite{Chen:2011}. Thus the Casimir repulsion found in \cite{Chen:2011,Chen:2012ds} for \emph{opposite} signs of the Hall conductivity actually corresponds to \emph{identical} signs. We trace this back to the use in \cite{Chen:2011,Chen:2012ds} of the Lifshitz formula with $\rb$ instead of $\rb'$ in TE-TM basis. However, since the conductivity (\ref{iii_sigmas}) satisfies Eq.\ (\ref{invEM}) this problem is easily curable.

Let us proceed with calculation of the Casimir force between two TIs separated by the distance $a$. To this end we utilize the  Lifshitz formula \Ref{Lif4} with reflection coefficients given by \Ref{R-expl} (and taking into account Eq. \eqref{rXY'}). The system consists of two TIs with parallel surfaces separated by a vacuum gap, see Fig. \ref{fig:3n}(b). The surface conductivity entering the reflection coefficients is given by \Ref{iii_sigmas} (continued into Euclidean signature with the help of Eqs. (\ref{Psi_E}) and (\ref{phi_E}), while for the dielectric properties in the bulk of TIs we assume that 
\be
\mu^\pm_{I,\II}=\epsilon^+_I=\epsilon^-_{\II}=1, \qquad  
\ep^+_{\II}(i\xi)=\ep^-_{I}(i\xi)  = 1 + \frac{g }{\xi^2 +\omega_R^2}
\ee 
for a simpler comparison with results of \cite{Chen:2011}. Here $g $ is the oscillator strength, $\omega_R$ is its frequency, and we neglected the contribution from possible damping. The lower index of the dielectric permittivity is the one inherent from the corresponding reflection coefficient, $\rb'_I$, $\rb^{\phantom{\prime}}_{\II}$.

\begin{figure}
	\includegraphics[width=8.2cm]{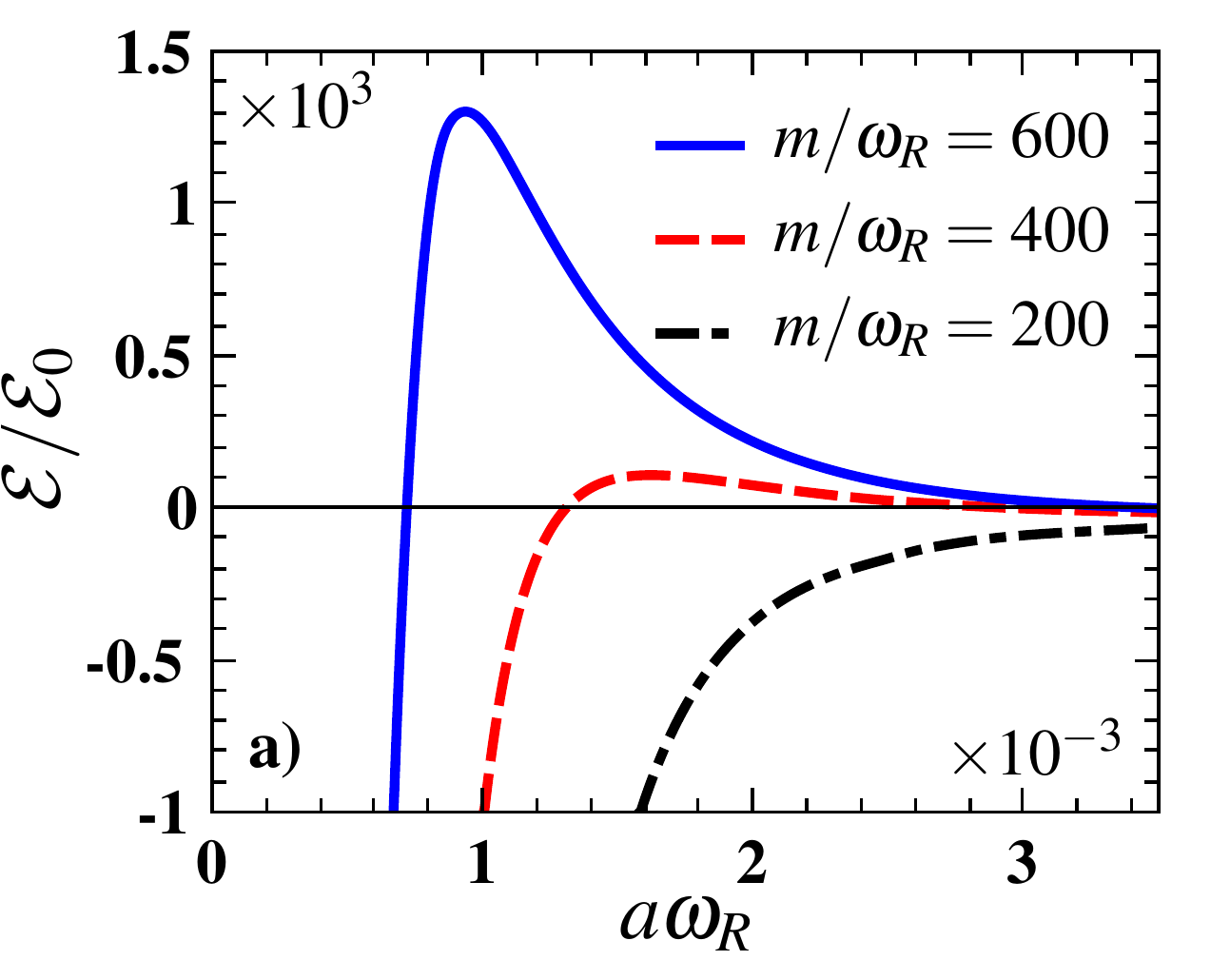}\includegraphics[width=8cm]{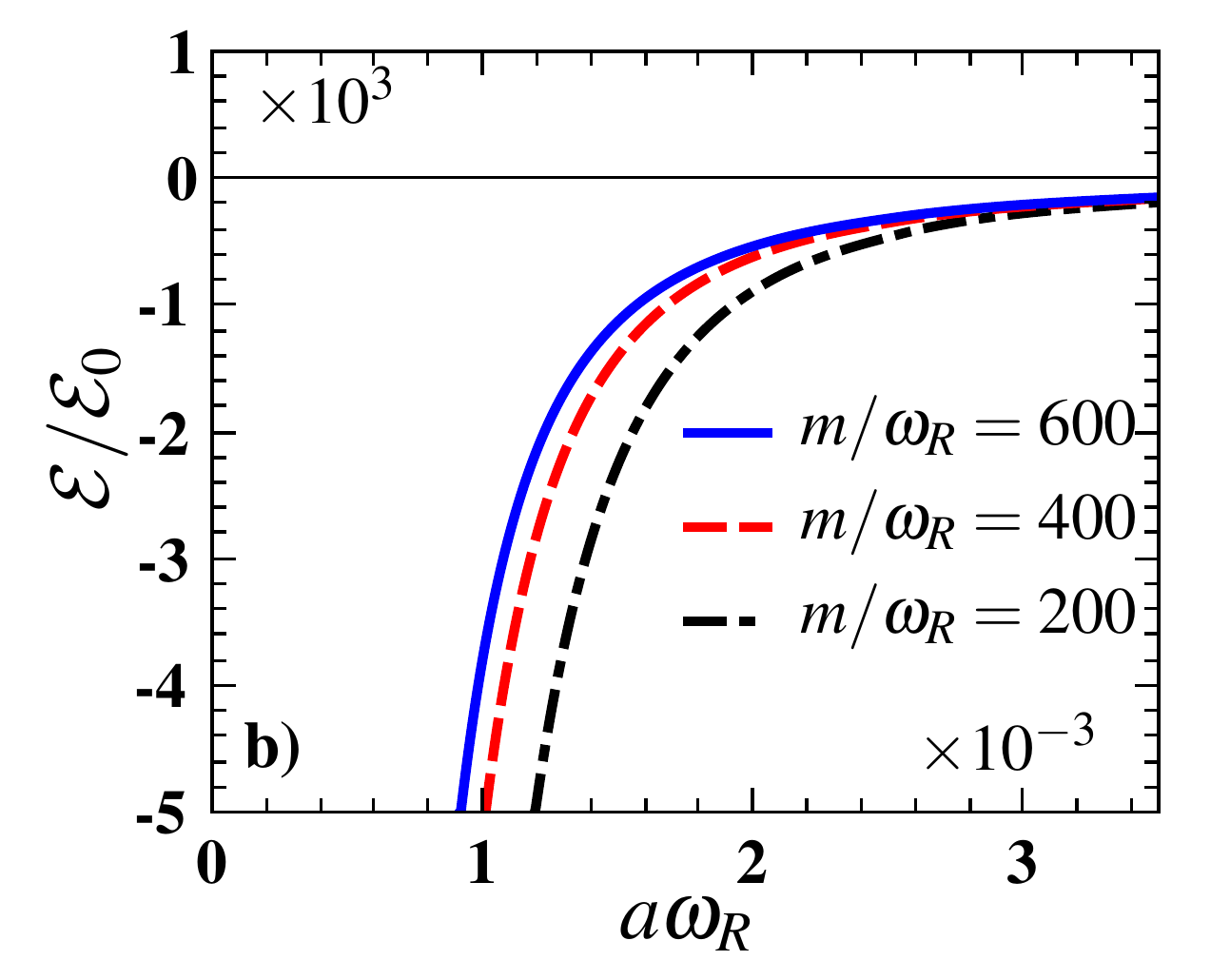}
	\caption{
	Casimir energy density $\cal E$ [in units of ${\cal E}_0 = \omega^3_R /(2\pi)^2$] as a function of the dimensionless distance $d =a\omega_R$  with different masses $m/\omega_R$ with anomaly contribution absent (panel a) and in presence of anomaly (panel b). Here we take the dimensionless oscillator strength $g /\omega^2_R = 0.45^2$, and Fermi velocity $v_F = 1.0\times10^{-3}$.}
	\label{fig:vs Chen}
\end{figure}

We performed numerical evaluation of the Casimir interaction in this case. Because of the presence of an additional dimensional parameter in the system, the oscillator frequency, we shall follow \cite{Chen:2011} and measure masses and distances in units of $\omega_R$, with $g/\omega^2_R = 0.45^2=0.2025$. Note, that the dielectric permittivity appears to be very close 1. To estimate the orders of relevant physical quantities, we need some value of the resonant frequency $\omega_R$. The paper \cite{Grushin:2011b} used $\omega_R \sim 56{\rm cm}^{-1}\sim 10^{-3}{\rm eV}$. Then $a\omega_R  = 10^{-3}$ corresponds to the distance $a\sim 180{\rm nm}$, which is a quite reasonable setup for Casimir experiments. 
As it is clear from the results presented on the Fig. \ref{fig:vs Chen} (where we followed as much as possible the notation and parameters values of \cite{Chen:2011}) the presence of anomaly completely removes any regions of possible Casimir repulsion in the system. 

Indeed, on the panel a) of Fig. \ref{fig:vs Chen} we plotted the energy for the same parameters' values as on  Fig. 4 of \cite{Chen:2011} in absence of anomaly.  The positive values of the energy (normalized to ${\cal E}_0 = \omega^3_R /(2\pi)^2$) corresponds to repulsion in the system, which is completely disappears as soon as one ``switches on'' the anomaly on the panel b) of the figure.

It is quite natural to expect. The conductivities used for plots on the left panel of Fig. \ref{fig:vs Chen} are those of (\ref{iii_sigmas})--(\ref{phi_E}) with ${\mathds A}=0$. In this case, the diagonal conductivities, $\si_{11,22}$, go to zero with large (infinite) masses, while the Hall contribution \Ref{phi_E} tends to a constant thus defining completely the interaction. However, we find it quite unnatural  if infinite mass quasi-particles give a non vanishing contributions to the  conductivity. 

In the opposite case, when the anomaly is present, the relative contributions of the Hall conductivity  for large masses  is much smaller then of the diagonal one, and the repulsion invoked by the Hall contribution is suppressed by the attraction of the normal one.

\begin{figure}
	\includegraphics[width=9cm]{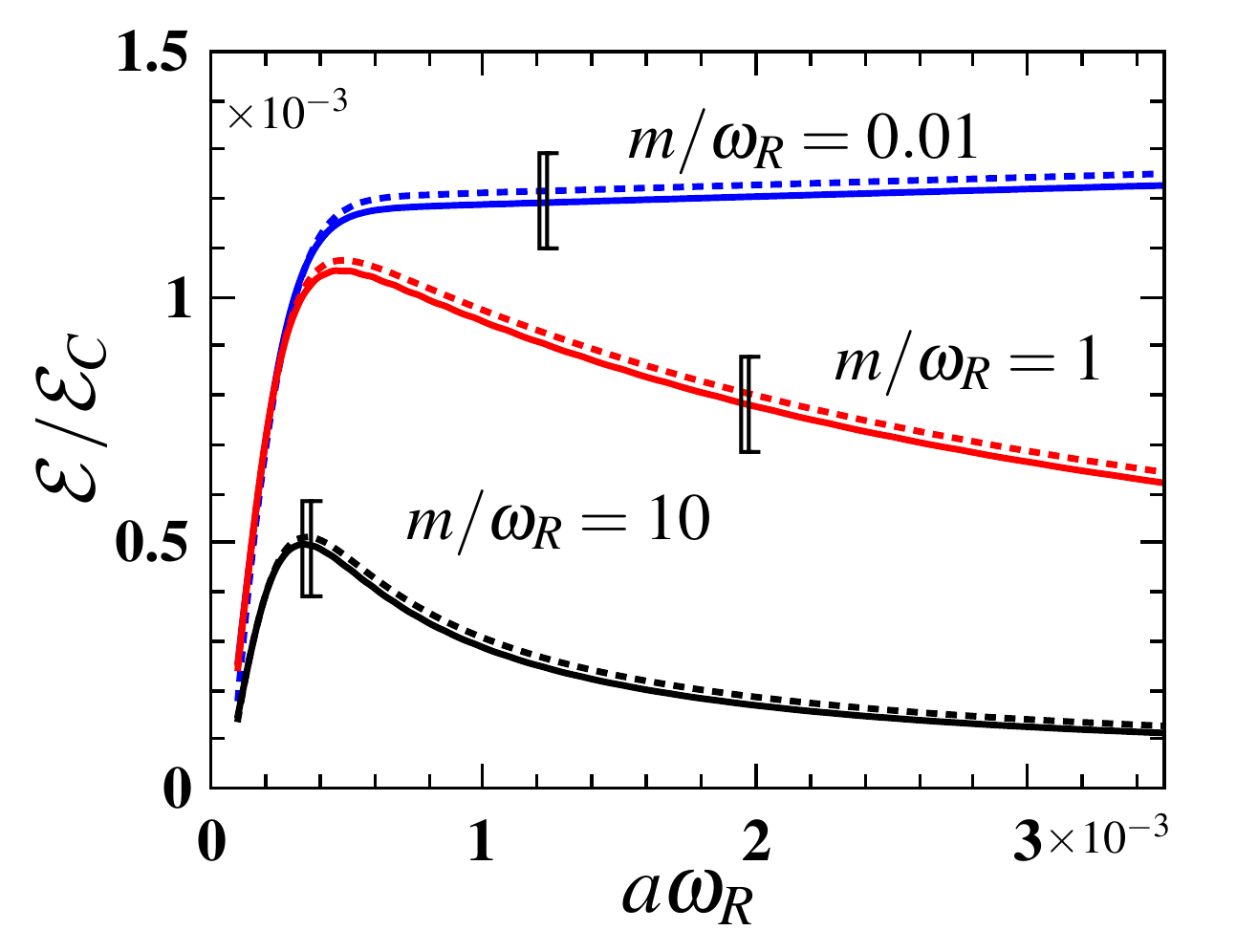}
	\caption{
	Ratio of the Casimir energy density between two identical TIs (full lines) and two opposite TIs (dashed lines) to two ideal metals, ${\cal E}/{\cal E}_C$ as a function of the dimensionless distance $d =a\omega_R$  with different masses $m/\omega_R$ in presence of anomaly. 
The dimensionless oscillator strength is $g_J/\omega^2_R = 0.45^2$, and Fermi velocity $v_F = 1.0\times10^{-3}$.}
\label{fig:small m}
\end{figure}

The values of mass parameter for Fig.\ \ref{fig:vs Chen} were taken in the same range as in \cite{Chen:2011}. For $\omega_R$ as presented above, $m$ appears to be about $0.2$ - $0.6$eV, which is quite close to the energy scale at which the continuum model is no longer valid. To cover the full range of reasonable mass parameters,
we present on Fig. \ref{fig:small m}  the energy of interaction of two TIs (in the presence of anomaly) for smaller masses both for identical Hall contributions (full lines) and opposite ones (dashed lines). The value of $m/\omega_R=0.01$ corresponds to practically massless quasiparticles. We use here a different normalization for the plot as compared to Fig. \ref{fig:vs Chen} and represent the ratio of the Casimir energy between two TIs to the one between two ideal metals, ${\cal E}/{\cal E}_C$. It gives a much better feeling of the energy values scale. For a growing mass the ratio decreases as a consequence of the decreasing conductivity of massive quasi-particles. All in all, the energy between two identical TIs is about $6$ times smaller then between two graphene layers. It is a consequence of presence of $N=4$ species of fermions in graphene, which at the level of expression \Ref{iii_sigmas}, \Ref{npnm} corresponds to $n^+=n^-=2$. As we can see from the plot, the presence of the Hall conductivity and its relative sign in two interacting interfaces is practically irrelevant. Only, a  week growth of the energy when switching from repulsive (symmetric) to attractive (antisymmetric) Hall configurations is visible.

Finally, a note is due on the relation between the fermion quasiparticles model for conductivity \Ref{iii_sigmas} and a much simpler representation of Section \ref{sec:part}. The latter is an approximation to the former with mass equal to zero, $m=0$, when the spatial dispersion is disregarded. This approximation is usually called the one of constant conductivity, and was investigated, for instance, in \cite{Khusnutdinov:2014:Cefswcc}. In this limit, the conductivities (\ref{iii_sigmas}) of single mode, $n^+=1$, $n^{-}=0$, give the parameters $\nxi=\alpha /32$, $\eta=\alpha/(4\pi)$. These values are extremely tiny. Looking at the graphs on Fig.~\ref{fig:1}, we see that all the lines start horizontally, showing practically no dependence on $\eta$. This reconfirms the observations made in this subsection.

\section{Conclusions}\label{sec:con}
In the first part of this paper we derived the Lifshitz formula for Casimir energy for generic planar (not necessarily isotropic) geometries. This derivation is suitable, in particular, for interacting surfaces with arbitrary conductivity tensors, and thus, for planar TIs. This derivation fills in an important gap in the literature and allows to explain some contradicting results on the sign of the Casimir force as a function of relative signs of Hall conductivities on the surfaces. Our conclusion is: the Casimir repulsion is only possible if the Hall conductivities on interacting surface have identical signs. We also derived a lot of potentially useful formulas regarding transformations between TE-TM and XY bases, reflection matrices for general conductivity tensors, etc.

As a particular model computation we studied the interaction of surfaces that have both constant longitudinal conductivity and a constant Hall conductivity. Among the most interesting findings is a critical value of longitudinal conductivity above that no repulsion is possible, and a quite pronounced dependence of the Casimir interaction of a surface with Hall conductivity to an ideal conductor on the value of Hall conductivity (though there is no repulsion in this configuration). Note, that this latter configuration is easier to realize at an experiment than two TIs. We also modified the model considered in \cite{Chen:2011} by adding a parity anomaly term to the conductivity. As we argued, this modification has very good physical grounds. With this term included, the Casimir force is always attractive.

Our work placed some additional limits on the Casimir repulsion in the Chern-Simons systems. This phenomenon is easily removed by enlarging the longitudinal surface conductivity. Nevertheless, there is still a window at large number of modes (or at large topological numbers). Another promising mechanism is the enhancement of Casimir repulsion in TI multilayers \cite{Zeng:2016}. The chemical potential may also increase the Casimir interaction, as was noted in \cite{Bordag:2015zda} in the case of graphene. The influence of chemical potential on the repulsion has not been studied so far with the only exception of \cite{Rodriguez-Lopez:2017:cfptitgf}. Although the temperature effect have been addressed already \cite{Grushin:2011,Chen:2012ds}, it is important to compute the combined affect of temperature and anomaly. We conclude by mentioning another interesting direction of reaserch, which is the Casimir-Polder interaction with Chern-Simons surfaces \cite{Marachevsky:2009eq,MartinRuiz:2018}. 

\begin{acknowledgments}
We thank Michael Bordag for long-time collaboration on the Casimir effect and related issues.
One of us (D.V.) is grateful to Valery Marachevsky for fruitful discussions. 
This work was supported in parts by the grants 2016/03319-6 and 2017/50294-1 of the S\~ao Paulo Research Foundation (FAPESP),  by the grant 303807/2016-4 of CNPq, by the RFBR projects 18-02-00149-a, 16-02-00415-a and by the Tomsk State University Competitiveness Improvement Program.

\end{acknowledgments}

%


\end{document}